\documentclass[]{elsarticle}
\usepackage{lineno,hyperref}
\modulolinenumbers[5]
\usepackage{amsmath}
\usepackage{graphicx} 
\usepackage{caption}
\usepackage{subcaption}
\DeclareGraphicsExtensions{.pdf,.png,.jpg,.eps,.ppm}
\journal{Journal of \LaTeX\ Templates}
\usepackage{color}
\usepackage{amssymb}
\usepackage{graphics}
\usepackage{epsfig,epstopdf}
\usepackage{epsfig}

\biboptions{sort&compress}

\begin{document}

\begin{frontmatter}

\title{Challenges of Numerical Simulation of\\ Dynamic Wetting Phenomena: A Review}

\author{Shahriar Afkhami}
\address{Department of Mathematical Sciences, New Jersey Institute of Technology, Newark, NJ, USA 07102}
\cortext[]{Corresponding author}
\ead{shahriar.afkhami@njit.edu}

\begin{abstract} 
Wetting is fundamental to many technological applications that
involve the motion of the fluid-fluid interface on a solid.
While static wetting is well understood in the context of 
thermodynamic equilibrium, dynamic wetting is more complicated
in that liquid interactions with a solid phase, possibly on 
molecular scales, can strongly influence the macroscopic scale
dynamics. The problem with continuum models of wetting
phenomena is then that they ought to be augmented with microscopic
models to describe the molecular neighborhood of the moving
contact line. In this review, widely used models
for the computation of wetting flows are summarized first,
followed by an overview of direct numerical simulations 
based on the Volume-of-Fluid approach. Recent developments
in the Volume-of-Fluid simulations of the wetting
are then reviewed, with a particular attention paid on 
combining macro-scale simulations with the hydrodynamic theory
near the moving contact line, as well as including a
microscopic description by coupling with the van der Waals
interface model. Finally, the extension to modeling 
the contact line motion on non-flat surfaces is surveyed, 
followed by hot topics in nucleate boiling.
\end{abstract}


\end{frontmatter}


\section{Introduction}
Wetting phenomena involves the motion of the interface between two or more fluids
in contact with a solid \cite{deGennesRMP}. Many natural and industrial processes 
are intimately related to the details of wetting phenomena. As technological capability 
evolves, it becomes increasingly important to understand the fundamentals of the fluid 
dynamical interactions involved in such applications, in order to optimize system performance. 
Numerical modeling offers great potential to predict outcomes under given conditions, 
and hence, ultimately, to optimize the process. Moreover, computational tools can play 
a very important role in terms of enhancing our understanding of the physics involved, 
and allowing predictions to be made cheaply.

Simulating wetting phenomena is complicated due to the presence of moving `contact
lines' (the lines at which a fluid-fluid interface meets a solid surface)
\cite{Jacqmin2000,Spelt2005,AB2008,afkhami_jcp09,afkhami_jcp2018}.
Considerable research effort has been directed to study such problem; see e.g.~\cite{Worner2012,Sui14}
for a comprehensive (and relatively recent) reviews. Simulating such phenomena is further complicated,
in part, by the fact that the description of moving contact lines involves widely varying characteristic 
lengths \cite{pismen00b,POMEAU2002,pismen00,STAROV1994}, from macroscopic to near molecular length scales. The motion of the 
fluid near the contact
line and the dynamic behavior of the interface itself present further complications, and analytical
solutions exist only for very simple problems. Numerical simulations can yield accurate predictions
but have been far less successful when studying wetting phenomena. For such problems, the efficacy of 
numerical simulation is highly dependent on the ability of accurately and robustly computing 
the moving contact line.

Another potential issue is that at the contact line, as the characteristic length
of the problem becomes smaller and the surface tension starts to dominate over other forces, 
 the surface tension discretization becomes a key ingredient in the feasible computation of such flows.
For these reasons, though reliable codes exist for the simpler two dimensional problem, there are
relatively few examples of three dimensional codes that incorporate algorithms for
treating dynamic contact lines; \cite{Bussmann99,Ding07,AB2009,Shin2018,yue_2020,LI2020,HAN2021} 
are representative of the variety of approaches that are
used. Moreover, robust and accurate numerical models are needed for simulating
fluid-structure interactions that involve moving solid objects in an interfacial flow
that also includes contact lines (see e.g.~\cite{OBrien2019,LI2020}).

While there exist different numerical approaches which can deal with dynamic contact lines,
here we concentrate on the `Volume-of-Fluid' (VOF) method, with our main goal to be to represent 
recent developments for the implementation of the contact angle boundary condition at a moving contact line,
self-consistent numerical validations, a dynamic contact angle model based on a hydrodynamic
description of the contact line, as well as an approach for the direct inclusion of liquid-solid interaction. 
We specifically pursue Gerris \cite{popinetGerris} numerical framework to describe
the implementations of the methods. The interested reader is referred to
the works in \cite{Popinet03,Popinet2009,PopinetARFM} for further details and discussions.
However, we also note possibilities of using other tools than Gerris, such as the use
of its successor code Basilisk \cite{popinetBasilisk}; see e.g.~\cite{Fullana2020,HAN2021,Sakakeeny2021} for very recent studies that utilize
Basilisk for the numerical simulation of dynamic contact line.
While the focus of this review is on a VOF method, we note that a major challenge 
for the simulation of moving contact lines in all continuum based approaches is to obtain mesh-independent results
with no adjustable parameters. The development of such models will require resolving all the length scales, 
typically from nanoscale at the solid boundary to bulk sizes, which can be of a millimeter size or more.
For feasible computations therefore sub-grid models ought to be introduced, and as a result, the
key ingredient in all current and previous models is to bridge the macroscale dynamics  
to the characteristics of the contact line at the nanoscale.

The nature of flows in a neighborhood of a moving contact line and the matching of local solutions to the
global flow have been the topic of a number of investigations; see e.g. 
\cite{Blake2006,Shikhmurzaev}. Despite the huge research effort, however, 
a full understanding of the mechanism by which a contact line moves along a solid
surface is still incomplete. While numerical investigations can enhance our understanding of the 
physics involved, simulating dynamic contact line flows is complicated by the mathematical paradox
of a contact line moving along a no-slip solid surface,
first discussed by Huh and Scriven in \cite{HuhScriv}. 
Here we will review major difficulties involved in continuum-level numerical 
simulations in the VOF numerical framework, mainly: (i) how to specify the dynamic
contact angle; and (ii) how to modify the no-slip boundary condition to remove
the stress singularity \cite{afkhami_jcp2018,Dziedzic2019}.

We first give a concise account of a variety of other computational methods 
that have been considered in the context of wetting phenomena.  
Here we mention level set methods that use a continuous function to describe the 
interface. A reconstruction of the continuous function in the ghost domain is used
to impose the contact angle within the level set approach \cite{LIU05,Spelt2005,ZHANG2020}.
(The grid points across the boundary are treated as ghost points).  
We also mention phase-field methods that treat two fluids
with a diffuse interface by means of a smooth concentration function, which
typically satisfies the Cahn-Hilliard or Allen--Cahn equations \cite{cahn_jcp58}, and is coupled
to the Navier--Stokes equations.  Jacqmin~\cite{Jacqmin1999} describes
a phase-field contact angle model that uses a wall energy to determine the value
of the normal derivative of the concentration on a solid substrate. This model
has been used to study contact line dynamics~\cite{Jacqmin2000,Jacqmin2004},
and similar models have been considered in the investigation of the sharp
interface limit of the diffuse interface model~\cite{Yue2010,Sibley2013}.
Lattice-Boltzmann methods have also treated the contact angle with a wall
energy contribution~\cite{Briant2004,Lee2010}. Molecular dynamics (MD)
simulations~\cite{Qian2003,Qian2006,Nguyen2012,fuentes_pre11} have also been considered, 
where the interaction between fluid and solid particles is described by
the Lennard--Jones potential, albeit mainly 
for simulating nanoscale systems.

In \cite{afkhami-kondic-2013}, the authors carry out continuum based computations
of dewetting of molten nanoparticles on a solid and show quantitative agreement with 
MD simulations, when either using a free-slip boundary condition, or
a partial slip with a slip length of nanometer sizes. 
In \cite{Ugis2020}, the authors carry out simulations of steady water 
drop sheared between two moving plates using MD, phase-field, and VOF methods. 
They show discrepancy, when comparing the results from phase-field to MD simulations. 
Moreover, they show that with a suitable choice of parameters and boundary conditions
in the phase-field simulations, a reasonable agreement with MD can be obtained. 
They also reproduce the MD results reasonably well with the VOF simulations
using a localized slip boundary condition. Interestingly, they point out the role
of diffusion in their phase-field model and the physical interpretation of it, showing
different effect to what happens in MD simulations, suggesting a perfect agreement
cannot be expected. In summary, the authors in \cite{Ugis2020} point out
the necessity for  hybrid  methods  for matching MD with continuum solvers,
similar in spirit to that in \cite{REN2005}.

Finally, in the context of front-tracking methods, in \cite{Huang2004},
the authors implement a front-tracking method and use a partial slip boundary 
condition and an ad hoc model for the determination of the dynamic contact angle
based on the contact line speed. 
Recently, in \cite{Shin2018}, the authors developed a front-tracking approach 
for simulating dynamic contact angles, where the contact line motion is
described by a slip model.

In what follows, we first present in Sec.~\ref{sec:vdW},
a multiscale wetting/dewetting model based on
an adaptive VOF method coupled with van der Waals interaction,
which can be considered as a mesoscopic approach
to model liquid-solid interaction. We then review a work on 
wetting transitions in Sec.~\ref{sec:MCL}. In Sec.~\ref{sec:IBM}, we discuss
an approach to model the contact line motion on non-flat surfaces and its
application to simulating a two-phase flow in porous media. Lastly, in Sec.~\ref{sec:ML},
as an example of a future direction, we present nucleate boiling, 
where resolving the details of the flow around a moving contact line is of a 
significant importance to building a generally applicable numerical model.

\section{A mesoscopic approach to model liquid-solid interaction}
\label{sec:vdW}
It is impossible for a contact line to move on a solid if a molecularly 
sharp fluid-fluid interface is considered along with a strictly no-slip solid.
The consequence of such is the divergence of shear stress upon approaching 
the contact line. As elegantly stated by Pomeau in \cite{POMEAU2002}, 
``One major difficulty met when trying to `solve' the moving contact line problem is to find at which
scale the usual continuum mechanics breaks down and should be amended to get rid of the divergence.''. Here, we review a recently developed computational scheme
for wetting/dewetting flows, that combines the van der Waals model with the
Navier-Stokes equations to devise a divergence free numerical framework for
the simulation of contact line motion \cite{MahadyvdW15}. We show
how this approach can lead to accurate description of the contact line,
while recovering the usual macroscopic scale flow far away from the contact line. 

Let us begin by considering the incompressible 
Navier--Stokes equations for a two-phase flow. 
We refer to two phases as the liquid phase (subscript $l$), 
and the vapor phase (subscript $v$), i.e.~$i=l,v$, with no loss of
generality. We introduce a characteristic function $\chi(x,y,z,t)$, 
which takes the value
of $1$ inside of the liquid phase, and $0$ inside the vapor phase. 
The interface between these two phases is assumed to be sharp, so that $\chi$
changes discontinuously at the interface.
The governing equations then are
\begin{equation}\label{eq:navierstokes}
\rho(\chi)\frac{D \mathbf{u}}{Dt} = -\nabla p + \nabla\cdot\left[ \mu (\chi)\left(\nabla
\mathbf{u} + \nabla\mathbf{u}^\top\right)\right]
+ \gamma\kappa\mathbf{n}\delta_s+ \mathbf{F}_{vdW}, 
\end{equation}
\begin{equation}\label{eq:incompressibility}
\nabla\cdot\textbf{u} = 0.
\end{equation}
Here, $\rho(\chi)=\rho_l\chi + \rho_v(1-\chi)$ is the density, 
$\mu(\chi)=\mu_l\chi + \mu_v(1-\chi)$ the viscosity, 
$p$ the pressure, 
and $\textbf{u}$ the velocity vector.
Also
\begin{equation}\label{eq:vof}
D\chi/Dt = \partial_t \chi + \mathbf{u}\cdot \nabla \chi =0.
\end{equation}
The surface tension is included as a singular body force (per unit volume)~\cite{Brackbill92,PopinetARFM},
where $\gamma$ is the surface tension coefficient, 
$\kappa$ the interfacial curvature,
$\delta_s$ a delta function centered on the interface, and $\mathbf{n}$ 
a normal vector for the interface pointing out of the liquid.
The total body force due to the van der Waals interaction
is represented by $\mathbf{F}_{vdW}$, which we describe next.

\begin{figure}
	\centering
	\includegraphics[height=3cm]{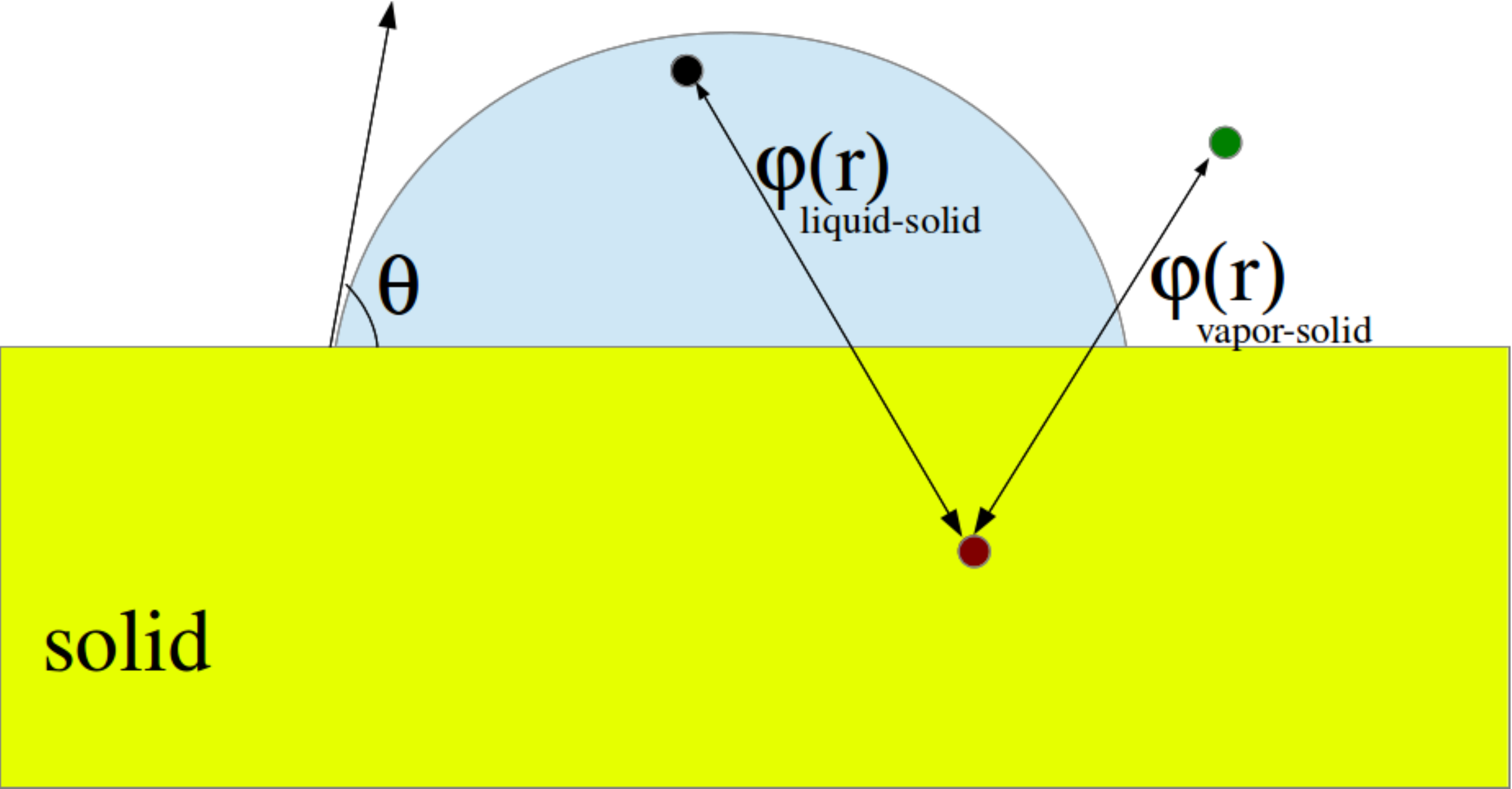}
	\caption{A schematic of the fluid-fluid-solid interactions.}
	\label{fig:1}
\end{figure}

We consider a solid phase occupying a half-infinite region
$y<0$, above which there is a region occupied by two fluids,
see Fig.~\ref{fig:1}.
Each particle of fluid phase $i$ interacts with the solid substrate by
means of a Lennard--Jones type potential~\cite{pismen00,STAROV1994}
\begin{equation}\label{eq:particlevdwpotential}
\phi_{is} = K_{is} \left( \left( \frac{\sigma}{r}\right)^a 
-\left( \frac{\sigma}{r}\right)^b \right),
\end{equation}
where $r$ is the distance between the two particles, and $K_{is}$ is
the scale of the potential. The total potential energy in phase $i$ due to this interaction is
\begin{equation}\label{eq:interaction}
\Phi_{is}(y) = \mathcal{K}_{is}\left[\left(\frac{h}{y}\right)^m -
\left(\frac{h}{y}\right)^n\right] = \mathcal{K}_{is} \psi(y),
\end{equation}
where
\begin{equation}\label{eq:strength}
\mathcal{K}_{is} = 2\pi n_i n_s K_{is}\sigma^3 \left(\frac{\left[(a-2)(a-3)\right]^{b-3}}{\left[(b-2)(b-3)\right]^{a-3}}\right)^{\frac{1}{a-b}},
\end{equation}
\begin{equation}\label{eq:equifilm}
h = \left[\frac{(a-2)(a-3)}{(b-2)(b-3)}\right]^\frac{1}{a-b}\sigma;\quad\quad  m = a-3, \quad n=b-3,
\end{equation}
where $n_i$ and $n_s$ are the densities of particles in fluid phase $i$ and  
the solid substrate, respectively. 
Equation \ref{eq:interaction} gives the total potential per unit volume in fluid
phase $i$ due to the interaction with the solid substrate.
The quantity $h$ is conventionally referred to as the equilibrium film 
thickness in the literature. The parameters $m$ and $n$ are 
taken based on the choices of $a$ and $b$ in Eq.~\ref{eq:particlevdwpotential}; see \cite{MahadyvdW15,MahadyvdW2015}, for example.

The force per unit volume on phase $i$ as a results of the potential,
Eq.~\ref{eq:interaction}, is $\mathbf{F}_{is}(y) = -\nabla \Phi_{is}(y)$, 
where
\begin{equation}\label{eq:total_body_force}
\mathbf{F}_{vdW} = \chi\hat{\textbf{F}}_{ls} + (1-\chi)\hat{\textbf{F}}_{vs},
\end{equation}
It can then be shown \cite{MahadyvdW15} that
\begin{equation}\label{eq:reduced}
\mathbf{F}_{vdW} = \nabla p_{vdW} + \frac{(1-\cos\theta_{\text{eq}})}{h}\frac{(m-1)(n-1)}{m-n}\,\psi(y)\mathbf{n} \delta_s.
\end{equation}
The first term on the right hand side of Eq.~\ref{eq:reduced}
is absorbed into the pressure gradient in Eq.~\ref{eq:navierstokes} for the entire domain, while the second term is centered only on the interface.
We refer the interested readers to \cite{PopinetARFM} for the details of the
discretization of $\kappa$ in Eq.~\ref{eq:navierstokes}, and $\mathbf{n}$
and $\delta_s$, in Eqs.~\ref{eq:navierstokes} and \ref{eq:reduced}. 

\begin{figure}
	\centering
	\includegraphics[height=6cm,trim=125mm 70mm 75mm 75mm, clip=true]{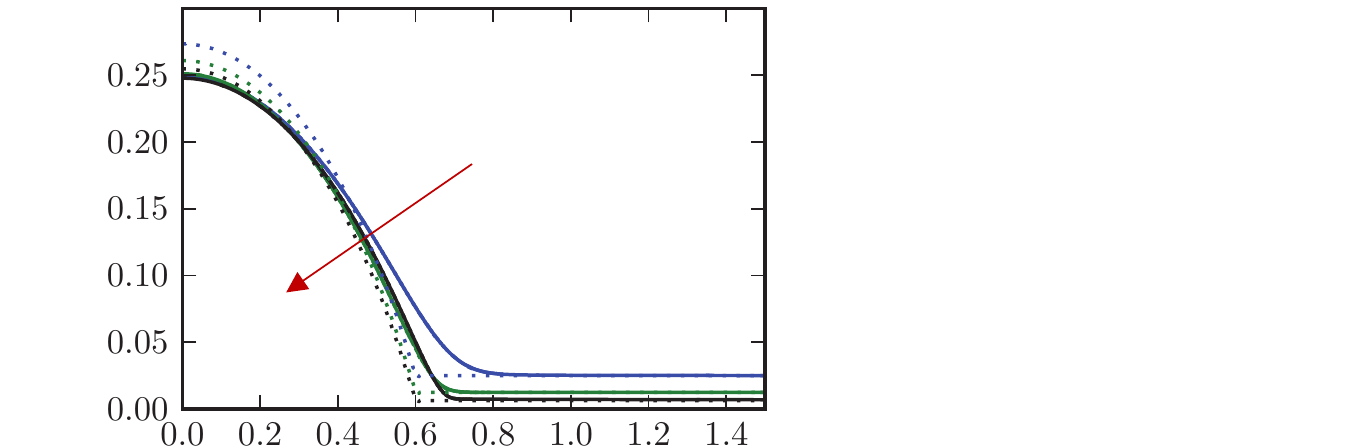}
	\caption{Equilibrium profiles (solid lines) 
		for $\theta_i = \theta_{\text{eq}} = \pi/4$;
		the arrow shows the direction of decreasing $h$.
		The dotted lines show the initial profile.
		$h=0.025$(blue), $0.0125$(green), $0.00625$(black);
		}
	\label{fig:h}
\end{figure}

The coefficient in the second term on the right hand side of 
Eq.~\ref{eq:reduced} involving the
equilibrium contact angle $\theta_{\text{eq}}$ can be derived based on simple 
energetic arguments (see \cite{MahadyvdW15}). In this context,
there is an equilibrium film of thickness $h$ that wets the 
entire substrate, with a smooth transition from the interface away from the 
contact to the film. The contact angle can then be defined by finding the
slope of the interface at the transition region. For example,
for a drop at equilibrium with a vanishingly small $h$,
we measure contact angles by fitting a circular profile
to the droplet profile `away' from the transition region; the angle at which
this circular profile intersects with the equilibrium film is taken to be the 
contact angle, $\theta_{\text{eq}}$.
For non-vanishing but small $h$, a drop at equilibrium will have a (slightly) different contact angle, which we refer to as $\theta_{\text{num}}$.
Figure \ref{fig:h} shows how the simulated equilibrium contact angle, $\theta_{\text{num}}$, is generally smaller than the
imposed angle $\theta_{\text{eq}}$.  This is due to the fact that
Eq.~\ref{eq:reduced} is derived under the assumption of small $h$.  
Figure \ref{fig:h} also shows the results when $h$ is varied from
$0.025$ to $0.00625$.  The initial condition is imposed with
$\theta_i=\pi/4$. As shown, the initial drops, depicted by
the black (dotted) profiles, relax to the equilibrium profiles, 
and as $h$ is decreased, the equilibrium profiles are 
characterized by contact angles closer to prescribed by 
$\theta_{\text{eq}}$.

\begin{figure}[th]
	\centering
	\begin{tabular}{cc}
		\includegraphics[scale=0.7]{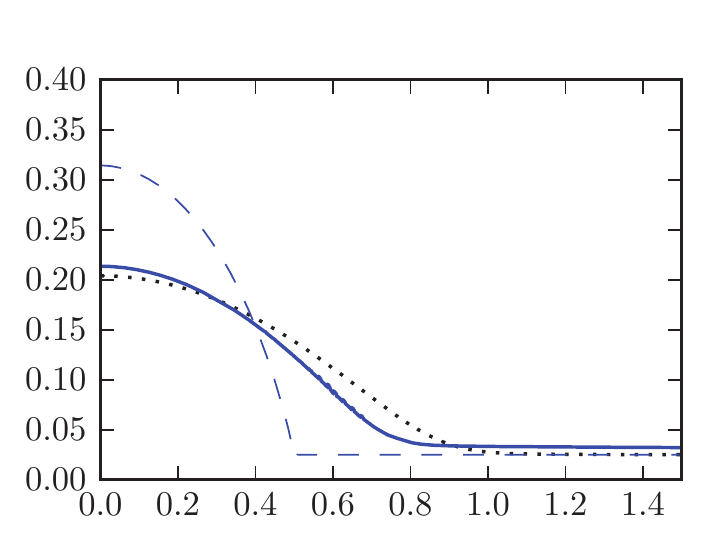}&
		\includegraphics[scale=0.7]{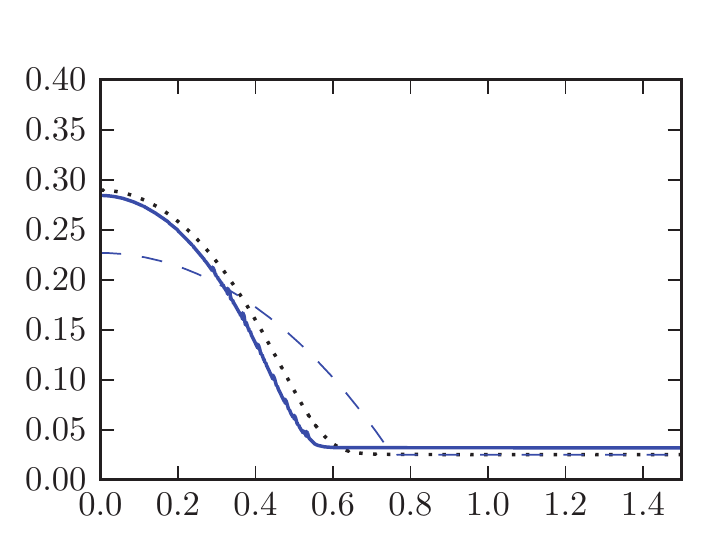}\\
		(a)&(b)\\
		\includegraphics[scale=0.7]{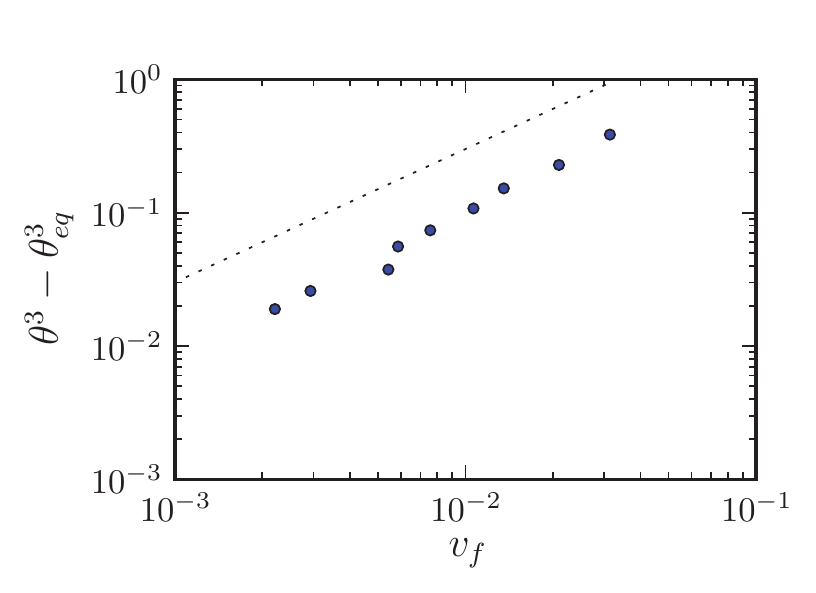}&
		\includegraphics[scale=0.7]{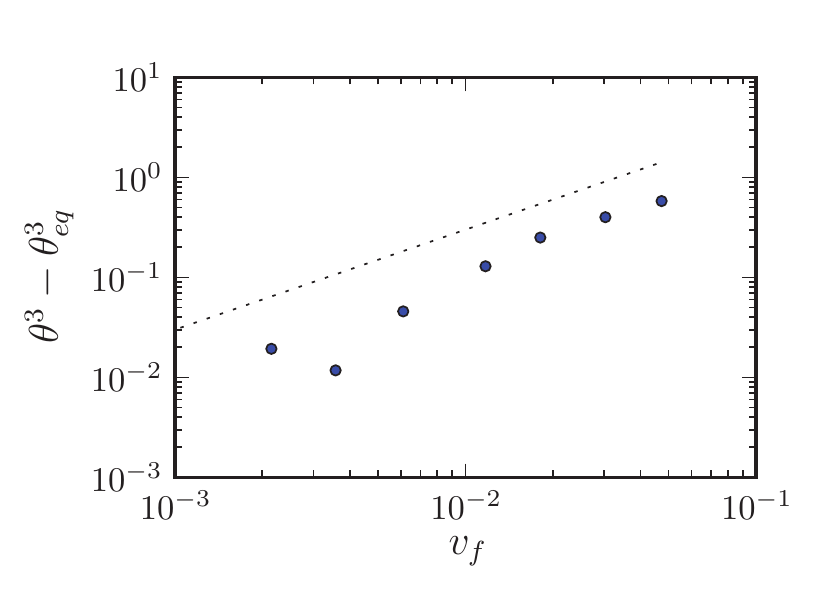}\\
		(c)&(d)     
	\end{tabular}    
	
	\caption{(a) Spreading of drop from initial contact angle
		$\theta_i=\pi/3$ to equilibrium defined by $\theta_{\text{eq}}=\pi/6$
		and (b) retraction from $\theta_i=\pi/6$ to $\theta_{\text{eq}}=\pi/3$. Comparison with the   	
		Cox-Voinov law for (c) spreading and (d) retracting drop. The blue (symbols) 
		shows simulation results and the black (dashed) line is
		proportional to $v_f$, showing the expected slope if the Cox-Voinov \cite{cox1986,voinov1976} law is obeyed. 
	}
	\label{fig:vdW1}
\end{figure} 
\subsection{Results}
We present the results of droplets with the contact angles imposed by means of
the discretization of Eq.~\ref{eq:reduced} as described in
\cite{MahadyvdW15,MahadyvdW2015}. The initial simulation setup is a drop on an equilibrium film, initially at rest,
which then relaxes to equilibrium under the influence of the van der Waals
force. For all the results shown next, $\rho_{l,v}=\mu_{l,v}=\gamma=1$, $h=0.025$, and $(m,n)=(3,2)$.
Figures \ref{fig:vdW1}(a-b) show the results of droplet spreading and retracting.
As shown, the droplet spreads/retracts to its equilibrium configuration, defined
by the $\theta_{\text{eq}}$. We also compare the qualitative behavior of the spreading/retracting drop to the well known
Cox-Voinov law~\cite{voinov1976}. For a drop displacing another immiscible fluid on a
solid surface, the speed of the contact point, $v_f$, is related to instantaneous angle $\theta$ and the equilibrium contact angle
$\theta_{\text{eq}}$ to the leading order by
$\theta^3-\theta_{\text{eq}}^3 \propto v_f$~\cite{cox1986}. As shown, the drop spreading/retracting follows the Cox-Voinov
law for $v_f \in (10^{-2},10^{-1})$.

\begin{figure}[t]
	\centering
		\includegraphics[scale=0.4,trim=0 110mm 0 0]{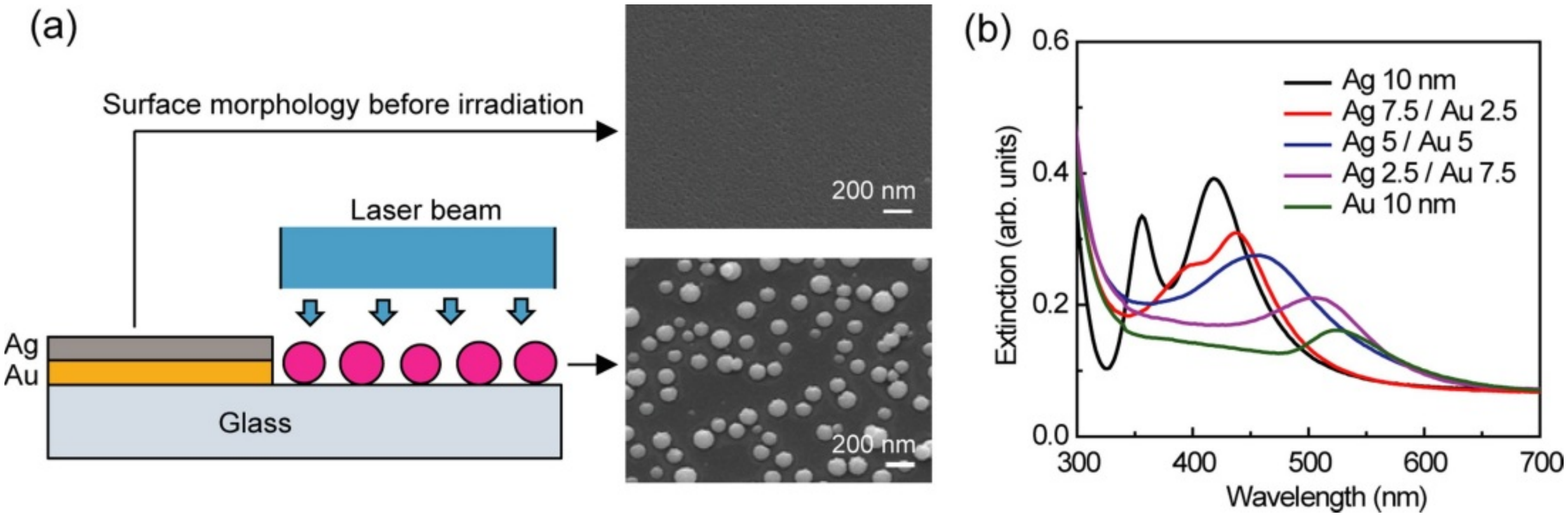}\\  	
	\caption{From \cite{OH2018}. (a) Morphologies of a bilayer thin film before and after laser irradiation. (b) Extinction spectra measured after thins films of different layer combination are dewetted. 
	}
	\label{fig:au-ag}
\end{figure}

Finally, we mention the dewetting phenomena of liquid films: for a sufficiently 
thin liquid layer (or multi-layers) on a solid substrate, destabilizing effects,
such as long range fluid-solid interactions, can lead to the film breakup 
and the consequent dewetting phenomena (exposure of prewetted/nominally dry patches), 
known as `spinodal dewetting' \cite{bonn_rmp2009}. Therefore, it is necessary to include
a destabilizing mechanism, such as the van der Waals interactions as described above, or
otherwise a continuous film does not breakup. While studying the dewetting of a thin
film is commonplace, see e.g.~\cite{seeman_jphys01,neto_jphys03,becker_nat03},
there is less effort in direct numerical simulations of multi-layer dewetting,
and the understanding of competing forces for such systems
is currently poor.  
Below, we describe our recent results of the dewetting of miscible liquid two-layers on 
substrates. In particular, we mention the work in \cite{OH2018} on the 
fabrication of Ag--Au bimetallic nanoparticles by laser-induced dewetting of bilayer films;
see for example, Fig.~\ref{fig:au-ag}.
\begin{figure}[t]
	\centering
	\includegraphics[scale=0.4,trim=0 20mm 0 0]{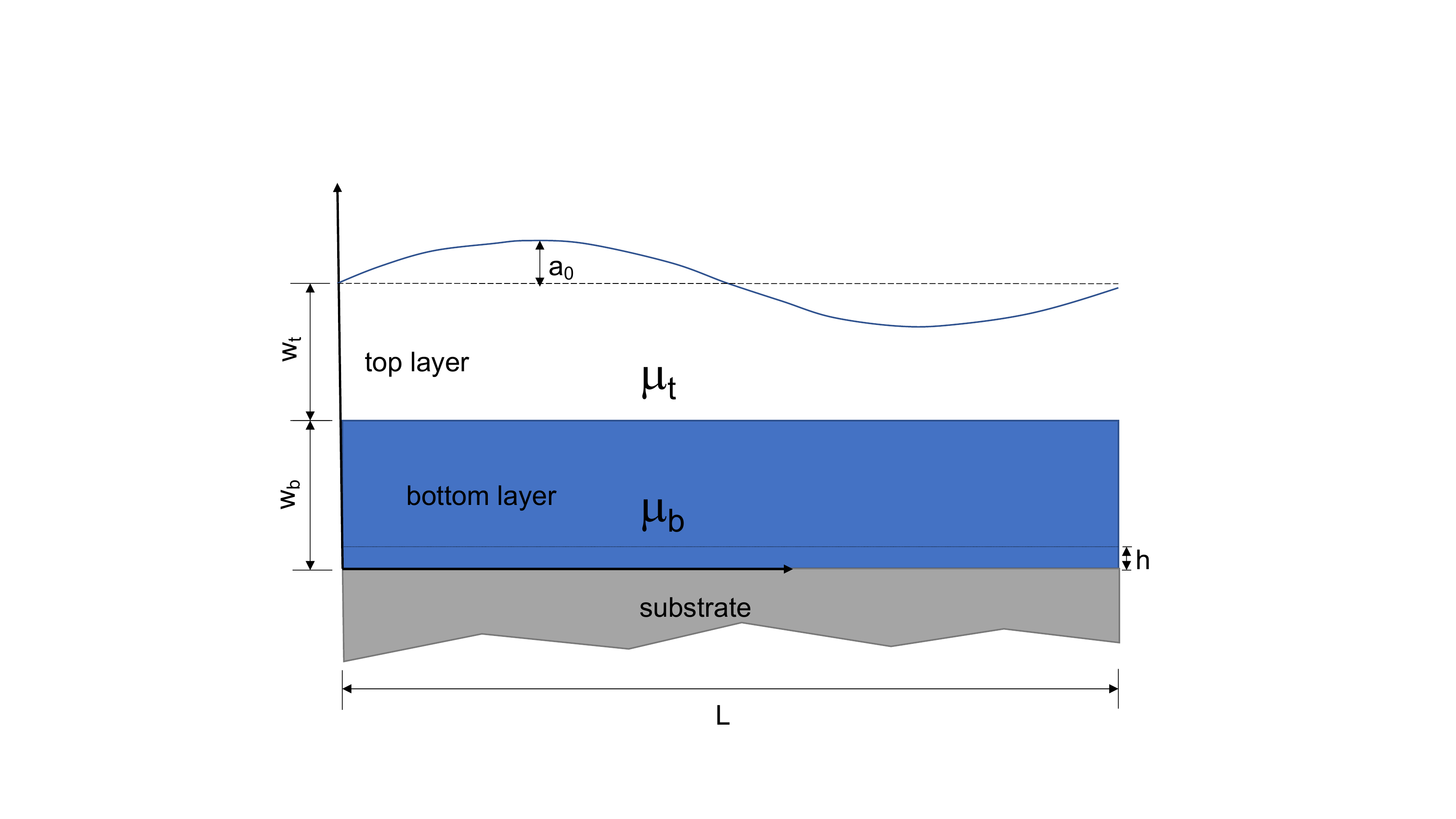}\\  	
	\caption{Schematic of the computational setup. Viscosity ratio is $\mu_b/\mu_t$,
		     $L$ is the wavelength of a linearly unstable sinusoidal wave with an amplitude
		     $a_0$, and $h$ is the prewetted or precursor film thickness.  
	}
	\label{fig:au-ag-sim}
\end{figure} 
\begin{figure}[th]
	\centering
	\begin{tabular}{cc}
		\includegraphics[scale=0.14]{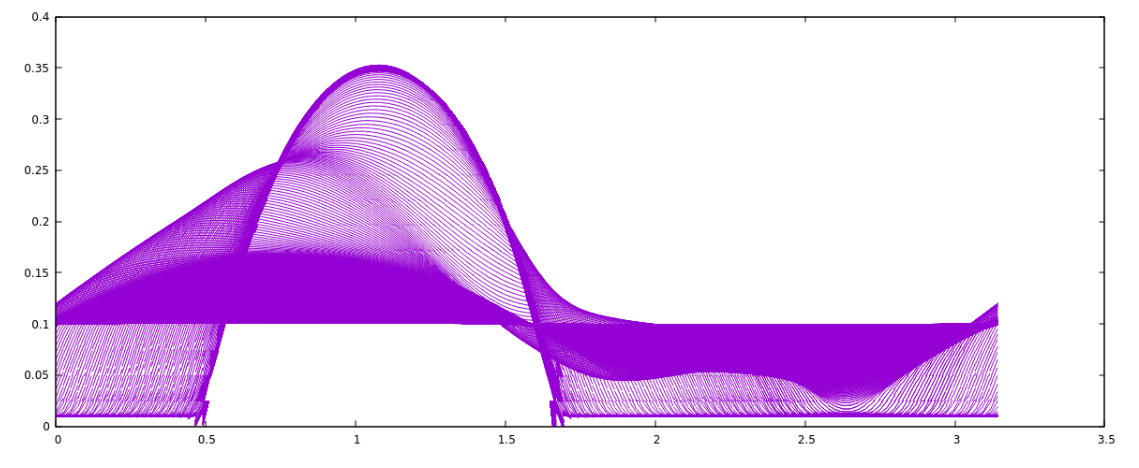}&
		\includegraphics[scale=0.14]{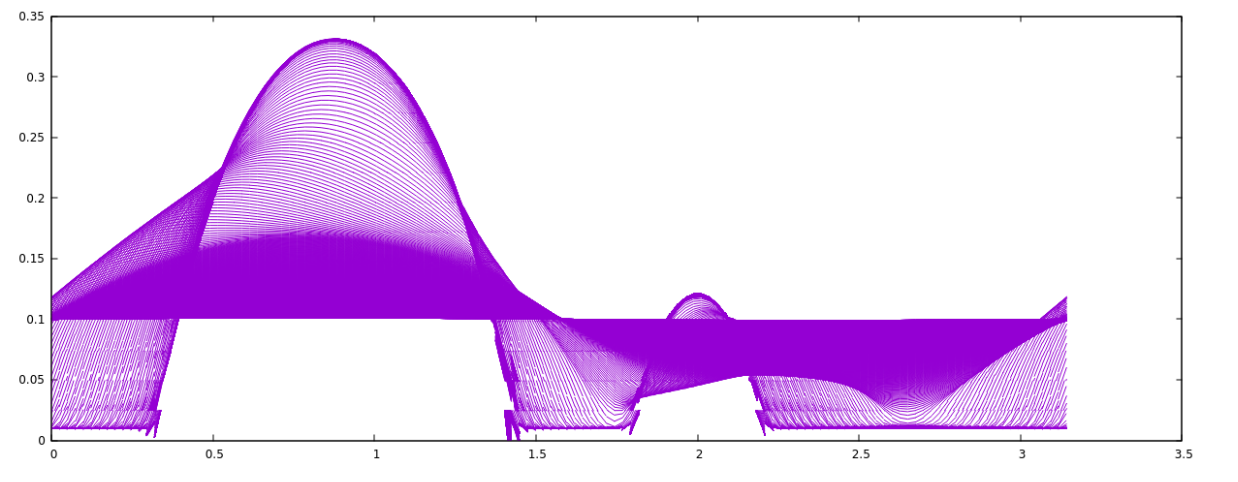}\\
		(a)&(b)\\
	\end{tabular}    
	
	\caption{Formation of a single drop (a) and two drops (b). $L=\pi$, $w_b=w_t=0.05$, $h=0.01$, $a_0=$ 1e-3, $\theta=90^\circ$, and (a) $\mu_b/\mu_t=1$  (b) $\mu_b/\mu_t=1.1$ . 
	}
	\label{fig:LSA}
\end{figure} 
Next we show examples of direct numerical simulations to study the structural evolution
and morphology of bilayer thin films consisted of miscible liquids. We will
concentrate in particular 
on the role of the contact angle and the viscosity ratio of the two layers
to study how changing these properties may affect the profile of the fluid as
it dewets. We keep the geometrical features fixed, and set the density ratio and surface tension to $1$, with a negligibly small surrounding viscosity. 
Figure \ref{fig:au-ag-sim} depicts 
the schematic of the computational setup. We then vary the contact angle and the liquid two-layer viscosity ratio
and consider two main scenarios where 
the breakup of the film results in either an unchanged length scale, namely leading to the formation of
one drop, or `dry spots', leading subsequently to the formation of one main drop and a secondary droplet.  The
findings here will help to understand how to control the synthesis of alloy metallic
nanoparticles, see e.g.~\cite{Rack2020}. Basilisk \cite{popinetBasilisk}, which is the Gerris successor
numerical framework, is used here.
We have specifically found: 1) In the absence of a second layer, the instability of the thin film follows
closely the predictions of the linear stability analysis \cite{MahadyvdW2015}; see Fig.~\ref{fig:LSA}(a). 2) By making the bottom 
layer slightly more viscous than the top layer, while the bilayer still remains unstable, 
the nonlinear instability
leads to the formation of a secondary droplet, suggesting that even a
small spatial variation of the viscosity can significantly alter the instability
characteristic length scales; see Fig.~\ref{fig:LSA}(b). 3)  There appears to be a correlation between the viscosity ratio and contact
angle defining the threshold for the formation of a secondary droplet; see Fig.~\ref{fig:cavr}.
\begin{figure}[th]
	\centering
	\includegraphics[scale=0.5,trim=20mm 30mm 0 0]{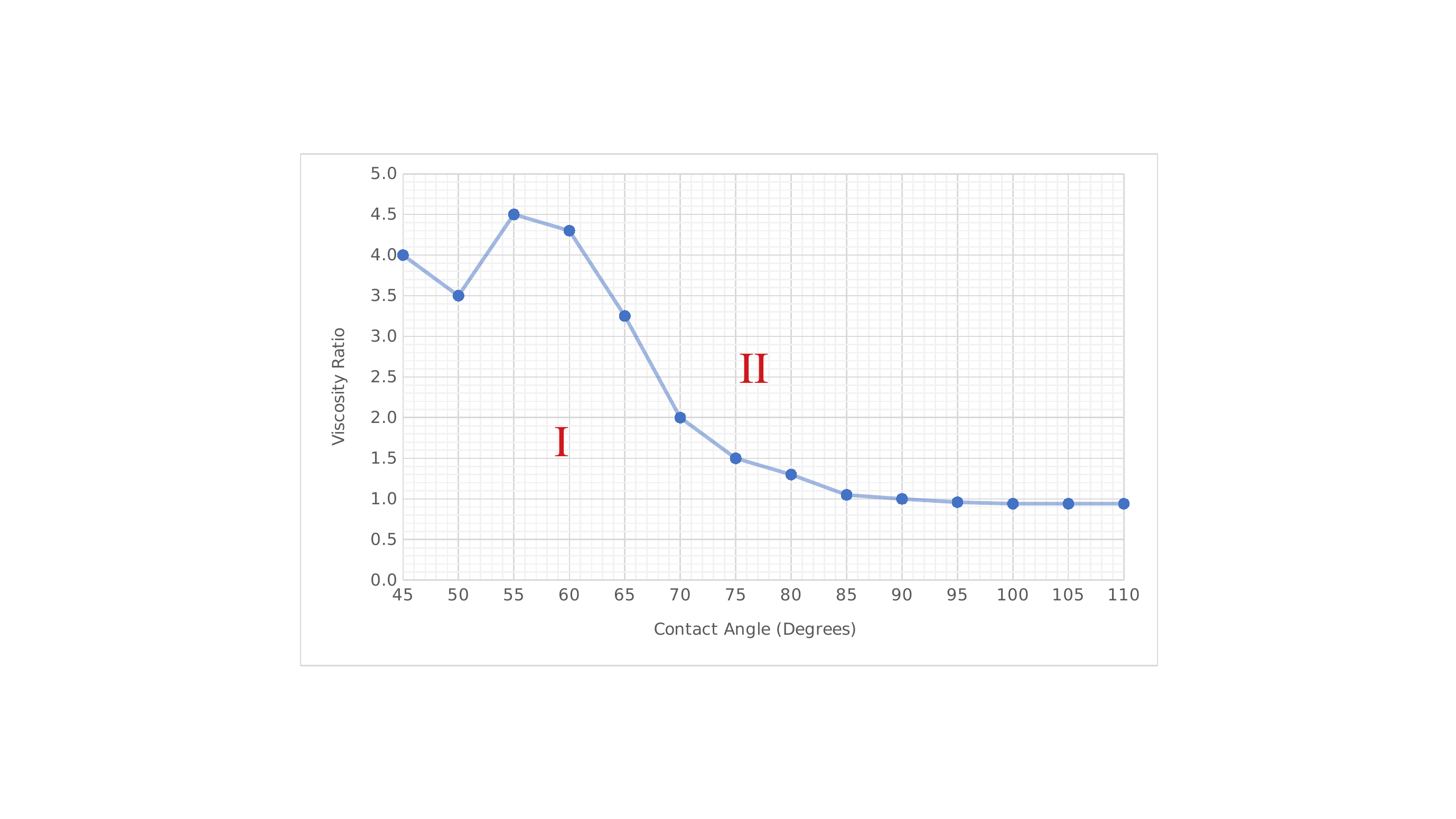}\\  	
	\caption{The threshold for the
		formation of the secondary droplet (symbols), showing the viscosity ratio $\mu_b/\mu_t$ as a function of the contact angle $\theta$.
		The geometrical parameters are the same as in Fig.~\ref{fig:LSA}. 
		Formation of a single drop ({\color{red}{I}}) and two drops ({\color{red}{II}}). 
	}
	\label{fig:cavr}
\end{figure} 

Our future work will involve investigating film lengths,
layer sequences, and the influence of the surrounding viscosity. We will
also further study the compositional structure of the bilayer systems.
Our preliminary numerical results reveal that the obtained final droplets have
a composition of the two layers over the whole volume of the droplets;
while our results show qualitatively similar distribution as illustrated in the experiments
in \cite{OH2018}, further analysis of the differences in the composition
as a function of the physical parameter set will help tuning
the morphology of such nanostructures for applications.

\section{A multiscale model for the simulation of dynamic contact lines: the dewetting transition}
\label{sec:MCL}

Here we review a recent work in \cite{afkhami_jcp2018} on the description
of a method for dynamic contact lines. The work focuses on in-depth 
analysis on the physical phenomenon of the dewetting transition.
When modeling dynamic contact line, here in the context of the VOF method,
there are two major challenges: 1) the numerical implementation of the contact
angle, and 2) relieving the incompatibility between the moving contact line
and the no-slip boundary condition (see e.g.~\cite{Fricke2020}, for a recent
mathematical view of the boundary conditions for dynamic wetting). 
In \cite{afkhami_jcp2018}, the authors
address both difficulties by first describing a method to impose the contact angle
as a boundary condition for when discretizing the surface tension in computational
cells adjacent to the solid boundary, and secondly, by devising an elaborate methodology to
account for microscopic features near the contact line, using an asymptotic theory
for the solution of the local contact line problem and to remove the
singularity in the shear stress, leading to numerical convergence.
We note however that a fundamental question still remains
regarding the measurement of the microscopic contact
angle and its definition at a dynamic contact line, see e.g.~\cite{AGRP}
for a recent and renewed discussion on the topic, and review papers \cite{bonn_rmp2009,Snoeijer13,Sui14}.

\newcommand{\be}{\begin{equation}}
\newcommand{\ee}{\end{equation}}
\newcommand{\nd}{\end{equation}}
\newcommand{\bea}{\begin{eqnarray}}
\newcommand{\eea}{\end{eqnarray}}
\newcommand\N{{\bf n}}
\renewcommand\Re{{\rm Re}\,}
\newcommand\Ca{{\rm Ca}}
\newcommand\Or{{\cal O}}

The discretization of Eq.~\ref{eq:vof} consists of two steps;
first the reconstruction of the interface followed by its advection. In the 
first part of the reconstruction step, the interface normal $\N=(n_x,n_y)$ in 
cel is determined from the values $\chi$ in neighboring cells, using
method described in \cite{AB2008,Popinet2009}. 
In the reconstruction step, the position of a linear segment 
representing the interface in the cell is determined 
from the knowledge of $\N$ and $\chi$ \cite{Tryggvason11}.
The reconstructed linear segments are updated in the
advection step, where the interface is evolved by the fluid velocity field,
using Eq.~\ref{eq:vof}.

In order to compute capillary forces, we use the Height-Function method, 
in which the local height of the interface is computed from summing over 
a column of cells \cite{AB2008}. Using second-order finite differences 
of the local Height-Function 
then provides the curvature, as well as the interface normals, used to compute
the surface tension force (see  \cite{Popinet2009,PopinetARFM}). 
Near the contact line, the interface is then linearly extended into the solid cell
to prescribe the value of $\chi$ function at ghost cells.
Having the interface Heights (and Widths) now defined 
at ghost cells, interface normal vector and
curvature adjacent to the solid boundary are computed as
usual. Once the interface positions and the curvature are computed, there is no 
special difficulty in computing the velocity field using the standard methods. 
No special provision is made for the discontinuity of velocities or the 
divergence of viscous stresses and pressures, which are computed as elsewhere 
in the domain using finite volumes and finite differences. 
Intuitively, this allows a kind of numerical slip. 
This numerical slip is studied in details in \cite{afkhami_jcp09}.

\subsection{Results}
\begin{figure}
	\centering
	\includegraphics[scale=0.25]{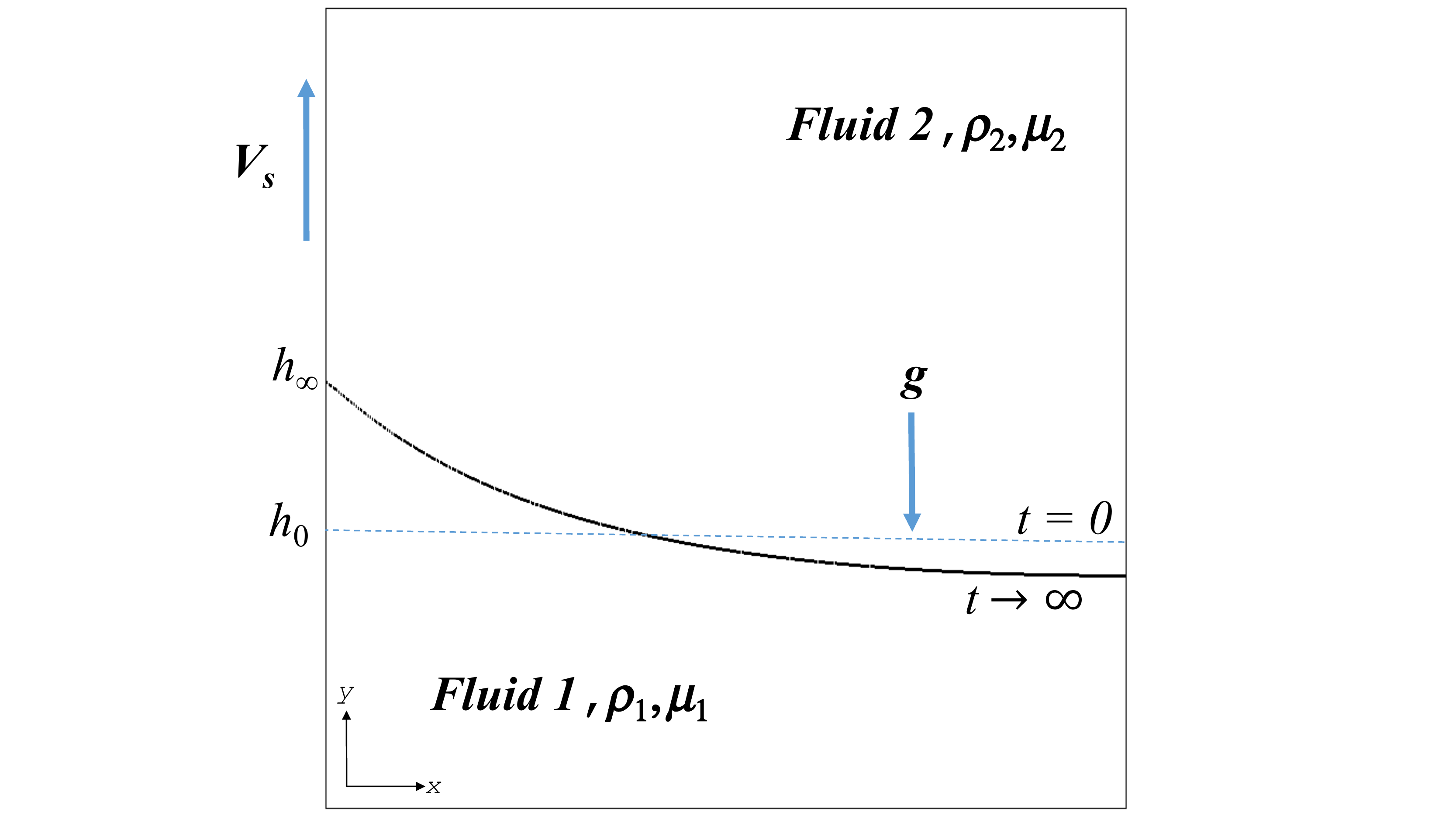}
	\caption{Schematic illustrating the withdrawing plate setup.
	}
	\label{fig:2}
\end{figure}
The problem setup is as follows: 
consider a solid plate being withdrawn from a liquid reservoir with
a constant velocity $V_s > 0$. The computational domain is $0\le x,y \le L$, 
with fluid~1 occupying $y<h_0$ and fluid~2 occupying $y>h_0$ at $t=0$ (see Fig.~\ref{fig:2}).
The viscosity and density of fluid $i=1,2$ are $\mu_i$ and $\rho_i$, respectively.
The capillary number is  defined as
$\mbox{Ca}=\mu_1 V_s/\sigma,$
where $\mu_i$ is the viscosity of fluid $i$ and
$\sigma$ the surface tension. 
We set $L\approx9\,l_c$, where $l_c$ is the capillary length
$l_c = \sqrt{\sigma/[(\rho_1 - \rho_2) g]}$
with $g$ the gravitational acceleration. 
The Reynolds number is then defined based on 
the capillary length as
$\mbox{Re} = \rho_1 V_s l_c/\mu_1$.

\begin{figure}[]
	\vspace{-10mm}
	\begin{center}
		\begin{tabular}{ccc}
			\includegraphics[width=2.5in, angle=90, trim=30mm 65mm 0 70mm]{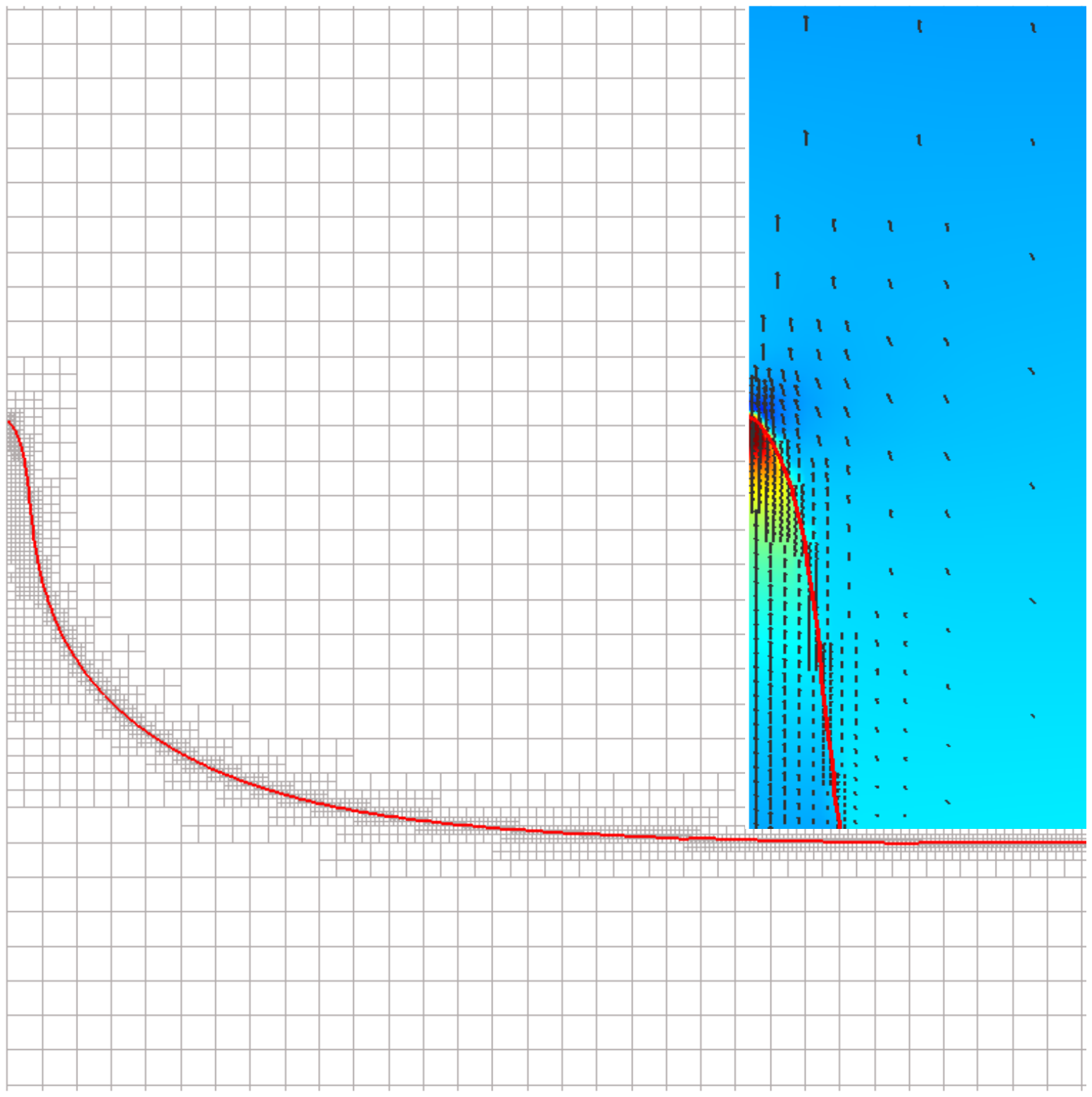}&
			\includegraphics[width=1.65in]{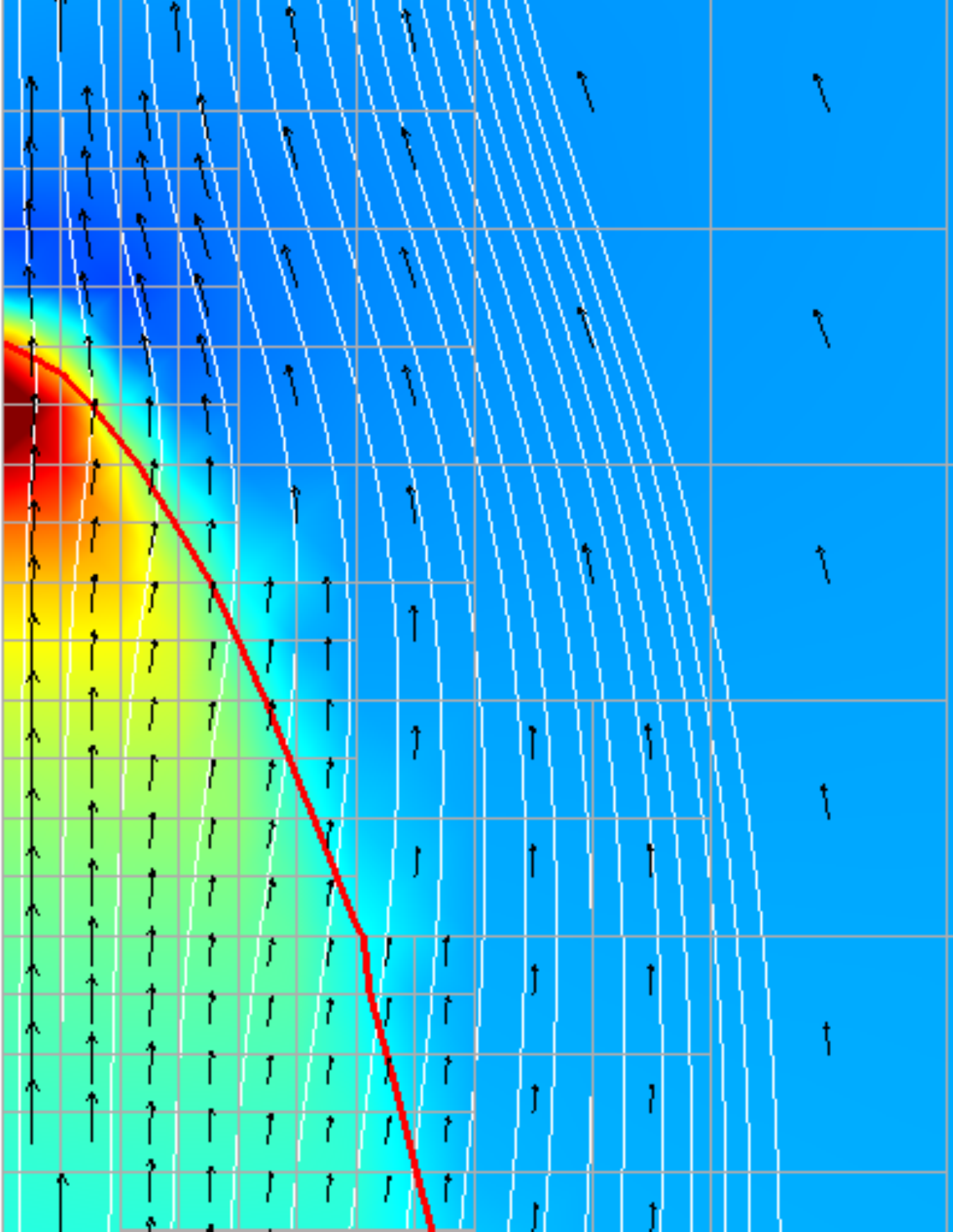}\\
			(a)\\
			\includegraphics[width=2.5in, angle=90, trim=30mm 65mm 0 70mm]{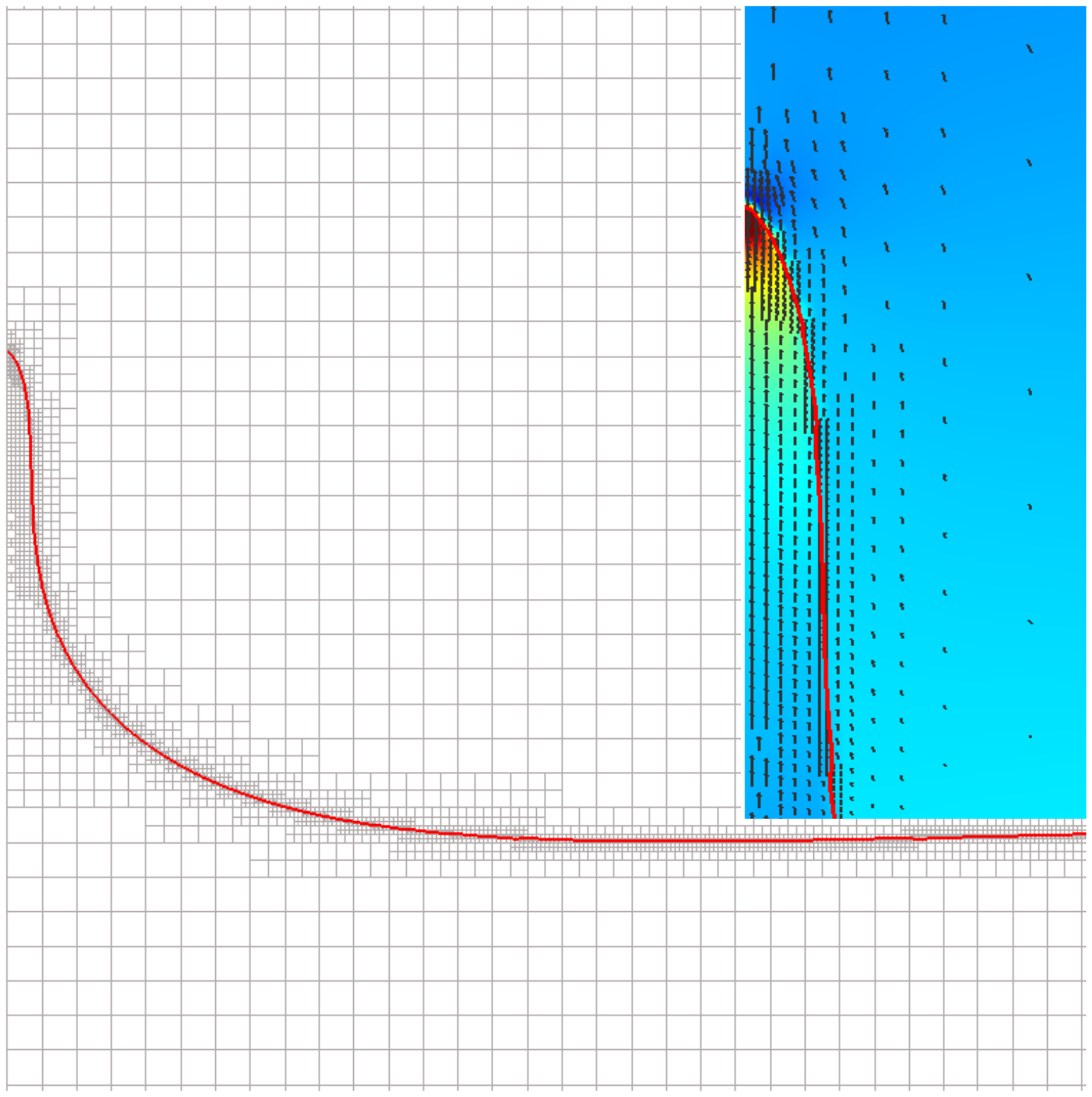}&
			\includegraphics[width=1.65in]{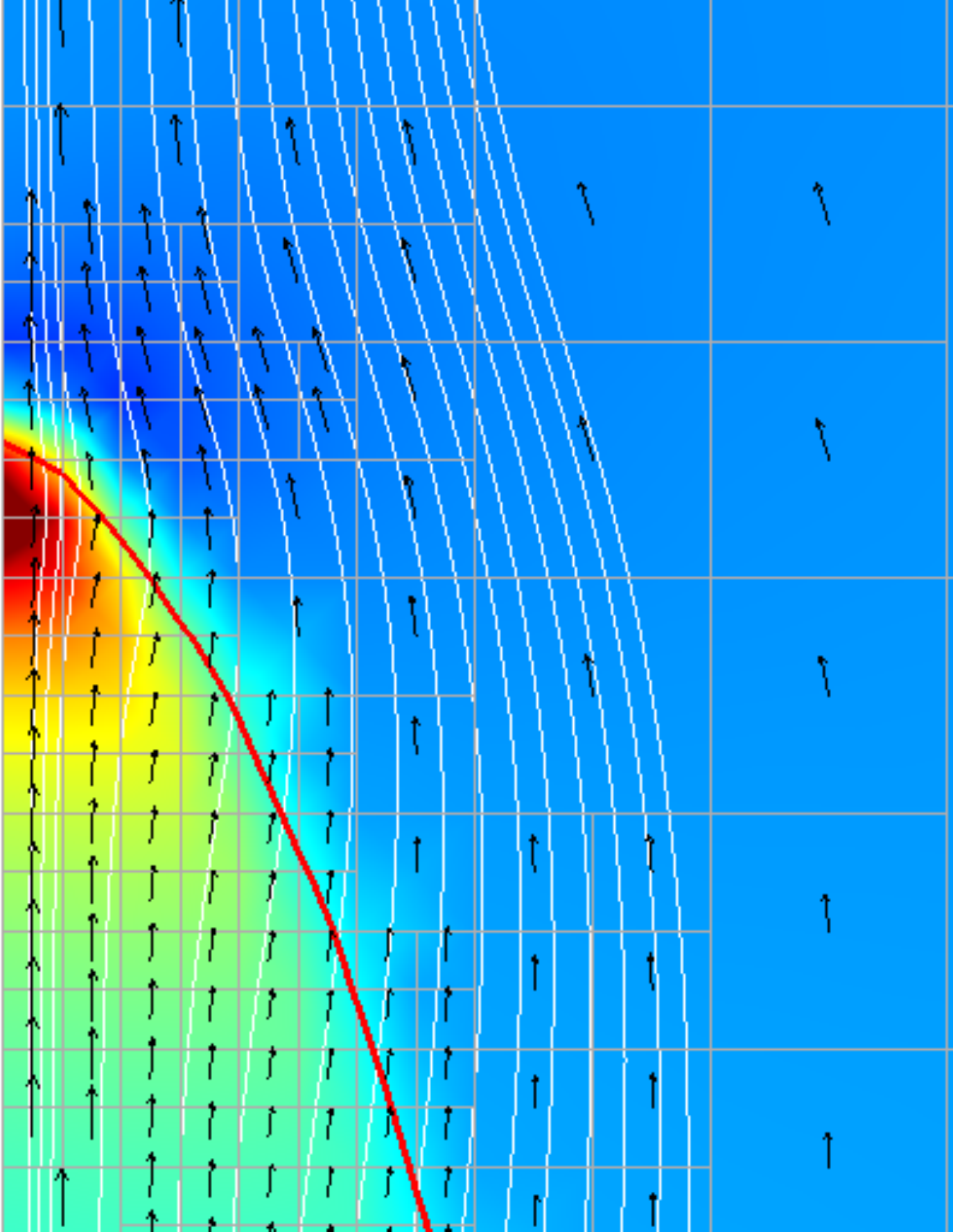}\\
			(b)\\
			\includegraphics[width=2.5in, angle=90, trim=30mm 65mm 0 70mm]{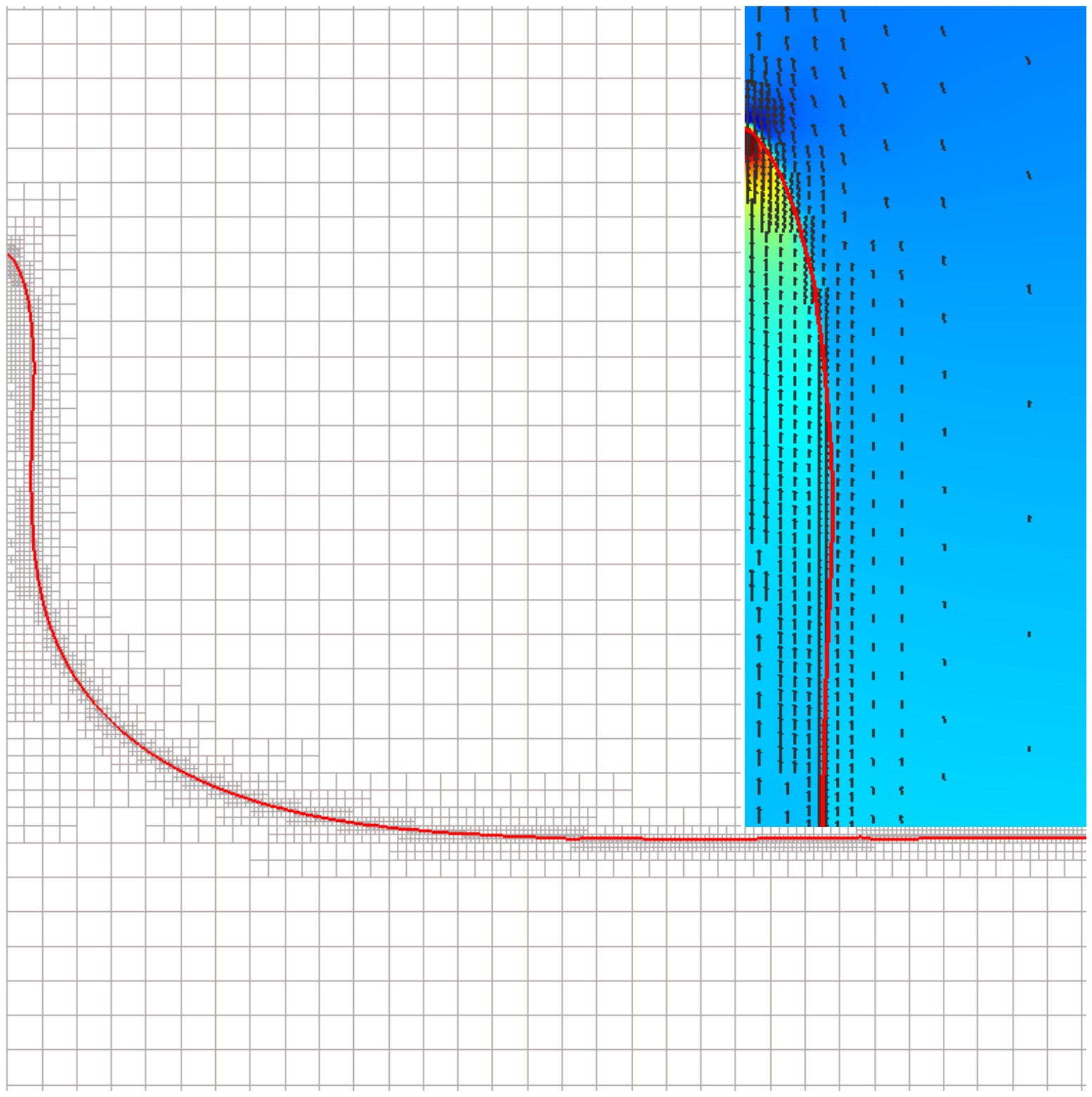}&
			\includegraphics[width=1.65in]{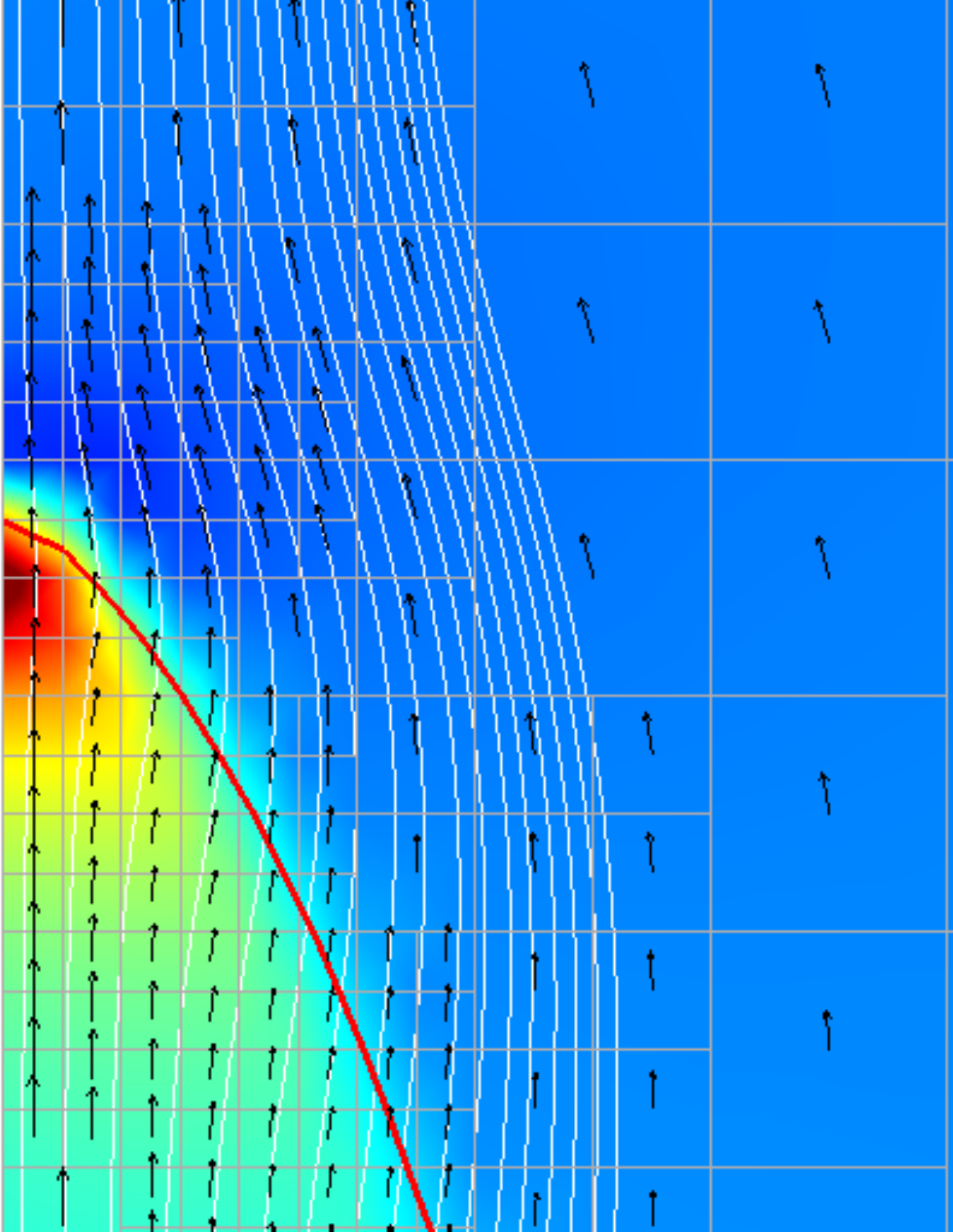}\\
			(c)
		\end{tabular}
	\end{center}
\end{figure}
\begin{figure}[]
	\begin{center}
		\begin{tabular}{c}
			\includegraphics[width=2.05in, trim=30mm 60mm 30mm 40mm,angle=90]{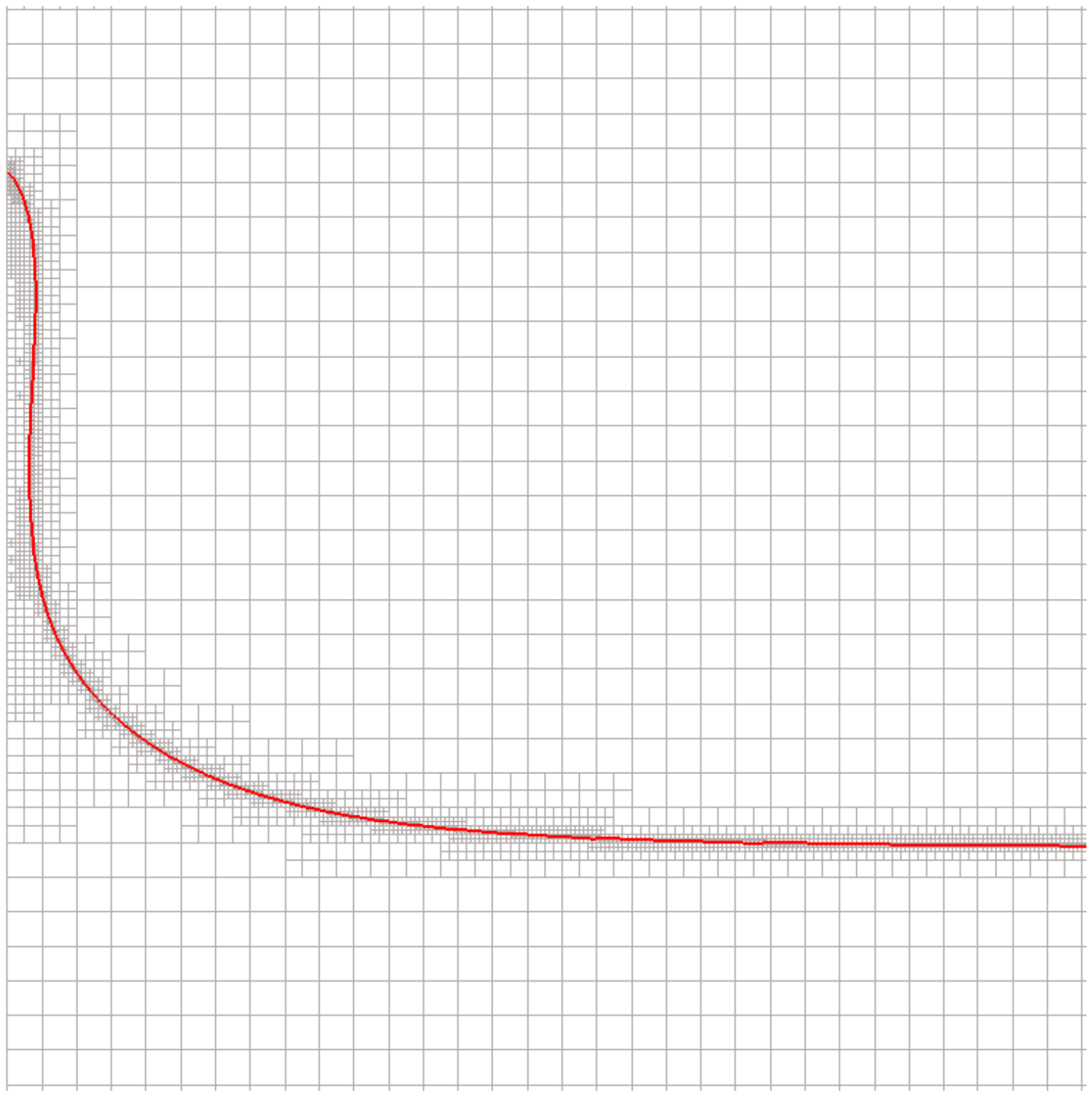}\\
			(d)\\
			\includegraphics[width=2.05in, trim=30mm 60mm 30mm 40mm,angle=90]{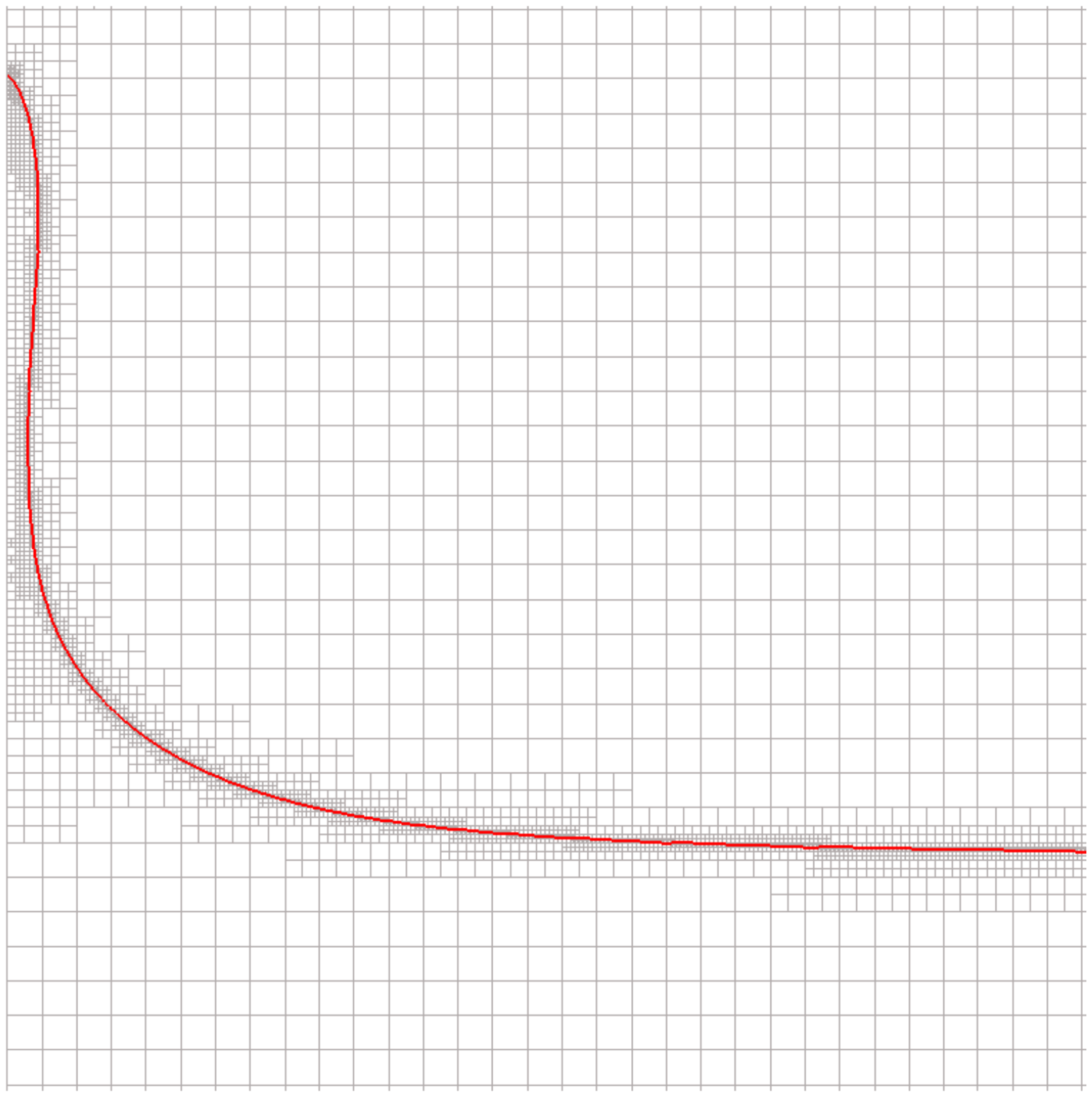}\\
			(e)\\
			\includegraphics[width=2.05in, trim=30mm 60mm 30mm 40mm,angle=90]{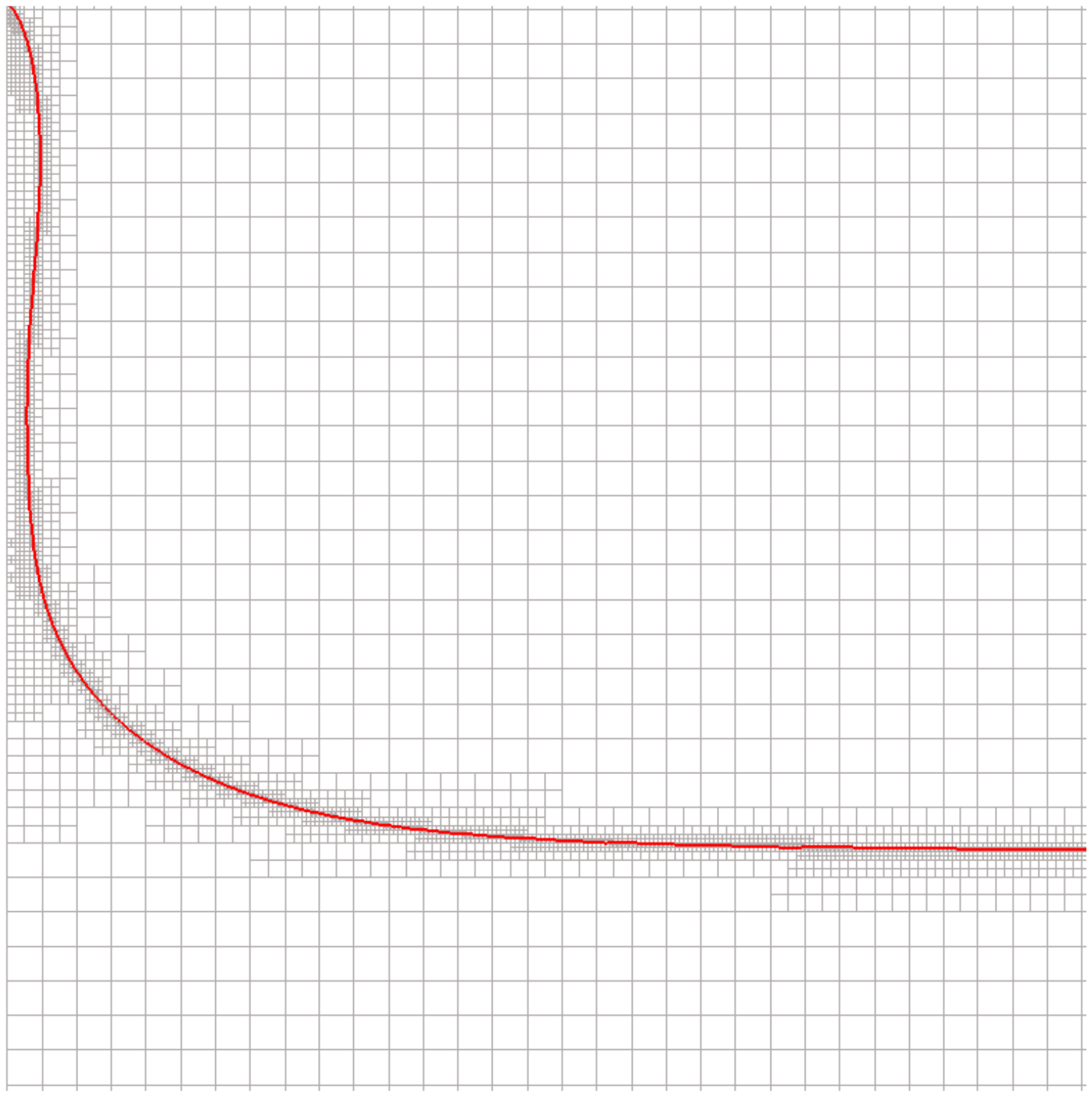}\\
			(f)
		\end{tabular}
	\end{center}
	\caption[]{From \cite{afkhami_jcp2018}. Time evolution of the interface for $\Ca>\mbox{Ca}_{cr}$ at
		$\tau = V_s t/l_c=$ 4 (a), 4.8 (b), 5.9 (c), 6.8 (d),  7.9 (e), 8.7 (f).
		The insets show the magnified flow field and the pressure distribution.
		(a)-(c) Right panels show a magnified view of the contact line region and the computational mesh;
		the fine structure of  the  flow field and the pressure distribution in the contact line region
		are illustrated. $\Ca=0.048$, and 
		$\theta=60^\circ$.
		The pressure colors show the maximum (dark red) and minimum (dark blue)
		of the pressure distribution.
	}
	\label{fig:simulation}
\end{figure}

In \cite{afkhami_jcp2018}, we focus on a specific, complex physical problem: the dewetting transition.
One example of such a flow is the withdrawing plate, 
see Fig.~\ref{fig:2}.
The interface may either sustain a stationary state meniscus,
if below a critical capillary number, ${\rm Ca}_{cr}$, or continue to move up 
the substrate until depositing a thin film to arbitrary heights.
The latter is called a Landau--Levich--Derjaguin (LLD) film \cite{LandauLevich1942,Derjaguin1943}.
Figure \ref{fig:simulation} from \cite{afkhami_jcp2018} shows an example of the
case where $\Ca>\mbox{Ca}_{cr}$.
In \cite{Eggers2004b,Eggers2005,Chan12}, 
a hydrodynamic theory is developed for the prediction 
of $\mbox{Ca}_{cr}$, while
Cox \cite{cox1986} and Voinov \cite{voinov1976} describe, 
on the other hand, how the singularity drives a peculiar
curved form of the fluid wedge at small $\Ca$.
In \cite{afkhami_jcp2018}, we use these theories to predict the numerically observed 
transition and to devise a subgrid scale model to reproduce 
the underlying microscopic physics. The latter reflects on the 
notion of grid-independent simulations in \cite{afkhami_jcp09,Legendre2015}.
In the simulations, we measure $\Ca_{cr}$ and specify $\Delta$,
the typical grid size, and $\theta_{\Delta}$, the specified 
contact angle at a particular grid size  $\Delta$. We compare the value of
$\Ca_{\rm cr}$ from full simulations to the solutions of the central relation
obtained in \cite{afkhami_jcp2018,afkhami_jcp2018_corr}
\begin{equation}
\frac{C(q) \phi(\theta_\Delta,q) \Ca_{cr}^{1/3} l_c}{\Delta} 
\exp \left[ - \frac {G(\theta_\Delta)}{\Ca_{cr}} \right] = 1, \label{cacr_cq}
\end{equation}
where the constant $C(q) = \Or(1)$, $q=\mu_2/\mu_1$, $G$ 
function is given in \cite{cox1986}, and $\phi = \Delta/r_m$
is the scaling factor or gauge function determined 
from the numerical simulations by comparing to the analytical solutions, 
where we consider $r_m \propto \Delta$, namely the numerical slip. 
In \cite{afkhami_jcp2018,afkhami_jcp2018_corr}, the computed $Ca_{cr}$
are in agreement with the above solutions.
Finally, it can be shown that 
\begin{equation}
\phi(\theta_\Delta,q) =
\frac{\pi{\mathrm e} {\rm Ai}^2(s_{\rm max})}{3^{1/3} 2^{-5/6} }
\frac {\Delta}{\Ca_{cr}^{1/3} l_c}
\exp \left[ \frac {G(\theta_\Delta,q)}{\Ca_{cr}} \right],
\label{phicomp}
\end{equation}
where $\rm Ai(x)$ is the Airy function of the first kind and
$\rm Ai^\prime(s_{\rm max})=0$; 
see  also an earlier work in \cite{Qin:18at}, where an expression similar to 
Eq.~\ref{phicomp} is derive.
 
\newcommand\solidrule[1][2mm]{\rule[0.5ex]{#1}{.9pt}}
\newcommand\dashedrule{\mbox{\solidrule[1mm]\hspace{1mm}\solidrule[1mm]}}
In Fig.~\ref{phifig}, we plot the values of the right hand side of  Eq.~\ref{phicomp}
along with the computed values of $\phi$.
In the near-free-surface Setup C ($q=1/50$),  a very clean estimate of the gauge function $\phi$
is obtained at small angles and we find 
$\phi(\theta_\Delta) \simeq  \theta_\Delta$. 
Thus our numerical model may be viewed as having an
effective slip of the order of a grid cell. However, we  find that there is a mismatch between the 
theory and the data in Eq.~\ref{phicomp} in Fig.~\ref{phifig}. If one
leaves the connection between the slip length $\lambda$ and the grid size $\Delta$ free, using
$\lambda = c \Delta$ with $c$ an arbitrary constant, then the adjustable constant $c$
allows a better fit of to the data; see, for example, the dashed line (${\color{black}\dashedrule}$)
in Fig.~\ref{phifig} for when choosing $c=3$. This in turn provides an estimate of
the effective slip in our VOF method.

\begin{figure}[t]
	\begin{center}
		\begin{tabular}{c}
			\includegraphics[width=4in]{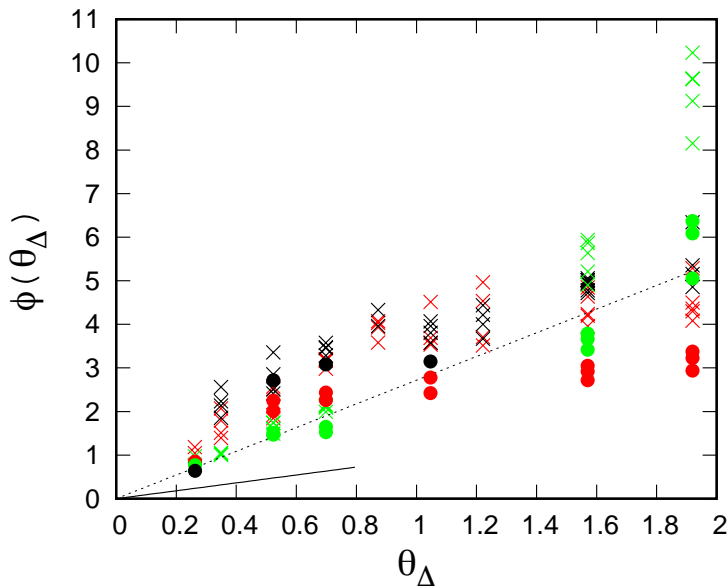}
		\end{tabular}
	\end{center}
	\caption{Using the data from \cite{afkhami_jcp2018,afkhami_jcp2018_corr},
	    $\phi$ is plotted using expression Eq.~\ref{phicomp} 
	    for Setups A: $q=1$ ({\color{black}$\times$}), B: $\Re=q=1$ ({\color{red}$\times$}), 
		and C: $q=1/50$ ({\color{green}$\times$}),
		compared to the computed values from the best fit of $\phi$ for 
		Setups A ({\color{black}$\bullet$}), B ({\color{red}$\bullet$}), and C ({\color{green}$\bullet$}), with $\theta_\Delta$ measured in radian.
		For a given angle $\theta_\Delta$, the various values of $\phi$ correspond to the 
		various values of the grid size $\Delta$ for each data set.
	    For all the simulations,
	    the density ratio $\rho_1/\rho_2=5$. 
	    The solid line ({\color{black}\solidrule}) is the prediction from lubrication theory  
	    for $\phi = {\mathrm e}\,\theta_\Delta /3$, when assuming $\lambda \approx \Delta$ and
	    dashed line ({\color{black}\dashedrule}), is for when adjusting  
	    $\lambda \approx 3\Delta$.
	    }
	\label{phifig}
\end{figure} 

\subsubsection{Recommendations for the numerical simulations of dynamic contact lines}
As we discussed above, numerical slip plays a crucial role in numerical simulations of the moving
contact line problem. As shown, this numerical slip, which is introduced at the
discrete level, is effectively equivalent to a slip length on the order 
of the grid size. Several authors have reported how this numerical slip can strongly distort 
the results \cite{Moriarty92,Weinstein08,afkhami_jcp09}, unless the grid size in the vicinity of the contact line is
decreased to a microscopic scale, making the computations prohibitively costly.  

To develop a strategy for realistic simulations where experiments would be used, 
the first step would be to perform a series of simulations at varying $\Delta$ and
for the conditions of the experiments. Obviously, simulations and experiments
are not performed near the critical $\Ca$ but below it. 
Comparing the angles observed in the region where the theory is still valid, would cross-validate the computations. This would in turn fix the parameters $\theta_{\Delta}$ and 
$r_m$ that could be used in simulations. 
This approach is of course made difficult by the fact that there
is no evidence so far that a single pair $\theta_{\Delta}, r_m$ 
could predict experiments over a range of 
different flow configurations. 
Actually the authors of \cite{YueFeng2011} state that ``In the literature we have 
not found a single pair of experiments in different geometries using precisely the same materials.''. 
This highlights
the difficulty of a predictive simulation approach based on experiments. Moreover, it should be
noted that the authors of references \cite{blake1999experimental} and \cite{Wilson:2006ks} 
claim precisely the opposite, that no microscopic angle, even a dynamical one depending on the capillary
number, can predict the whole range of experiments they have performed or simulated. 
We therefore believe that the relationship  between numerical  slip  length and  the  grid
size is strongly problem dependent. We provide such a relation in Eq.~\ref{phicomp} for the
withdrawing plate problem in the hope that it motivates further investigations for other scenarios,
such as droplet impact on a solid surface, where such relationships can also involve other
relevant dimensionless parameters such as Reynolds and Weber numbers. A suitable numerical slip
can then be identified by comparing the results with appropriate experimental measurements. 
We note however again that such a relation might only be applicable to that particular problem. 

Finally, the slip boundary condition combined with a constant contact angle might not sufficiently 
describe the dynamics of the moving contact line. In \cite{Fullana2020}, the authors propose to
replace the slip boundary condition with a generalized Navier boundary condition (GNBC) 
\cite{Qian03}, coupled with a dynamic contact angle to more favorably match the experimental observations
in the context of the curtain coating configuration \cite{blake1999experimental} (see also
\cite{FRICKE2019,Fricke2020} for a recent discussion on the boundary conditions for dynamic wetting).

\section{Moving contact lines on arbitrary geometries}
\label{sec:IBM}

Numerically modeling  non-ideal solid substrate is enormously complicated. 
Wetting and spreading on rough surfaces, two-phase flow over topographically patterned surfaces and through
porous media are examples of such systems. While there exist numerical studies on chemically disordered
surfaces, see e.g.~\cite{Wang2008,Xu2011}, and with roughness effects in a two-phase flow model, see
e.g.~\cite{Ren2014,LUO2017,Li2021}, pore-scale numerical investigations are still scarce, see
e.g.~\cite{Ferrari2013,Raeini2014,Bakhshian2019}, 
and therefor in what follows, we only focus on the direct
numerical simulation of fluid–fluid flows in a porous media model.

Pore-scale models have increasingly become a reliable tool for making predictions of 
two-phase flows through porous media;
see e.g.~\cite{Zhao2019,Ghillani2020}.
Direct numerical simulation techniques based on solving the full Navier-Stokes 
equations provide an accurate solution
for pore-scale analysis of multiphase  flows  in  porous  media.   
However, a great challenge in pore-scale direct numerical simulations is the inclusion of 
wetting effects into the  numerical model.
There exist direct numerical studies on pore-scale modeling of
wettability effects, such as the phase-field simulation in \cite{BASIRAT2017} and the lattice Boltzmann simulation in \cite{LIU2014}.  
While the majority of previous (Eulerian) interface capturing methods include wetting on body-fitted meshes, less progress has been 
made to develop numerical methods that handle arbitrary geometries (see e.g.~\cite{LIU05}). To that end, one attractive approach is 
the Immersed Boundary Method (IBM) \cite{Peskin1972,Mittal2005}, because of its ability to model complex geometries, 
and its potential for modeling systems of solid bodies with arbitrary relative motion
\cite{OBrien2019}. Here we show an accurate and robust numerical model developed in 
\cite{OBrien2018} and later extended in \cite{OBrien2019} for combining wetting dynamics, in a VOF framework, and the IBM. 
Here we briefly review the methodology developed in \cite{OBrien2018}
as the basis of an IBM for implementing wetting effects into the  numerical model
for arbitrary geometries.

\begin{figure}[]
	\begin{center}
			\includegraphics[width=4in]{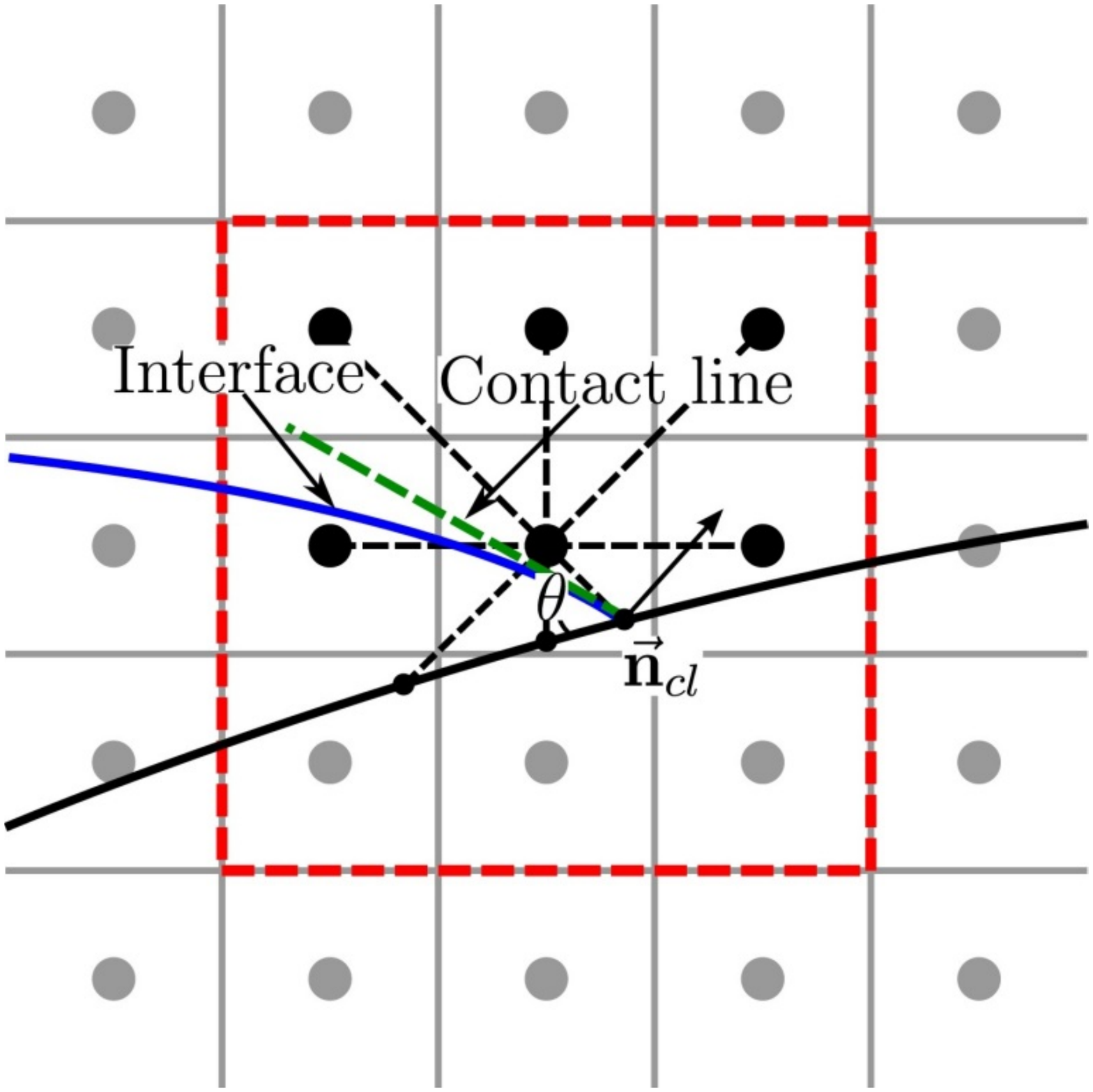}
	\end{center}
	\caption{From \cite{OBrien2018}. The stencil used for the evaluation of 
		curvature, $\kappa$, near the IB surface,
		the normal to the interface at the contact line is determined
		from the prescribed contact angle.}
	\label{fig:IB}
\end{figure} 
The method developed in \cite{OBrien2018} is based on a ghost-cell Immersed Boundary (IB) method for curved surfaces. In order to compute the interface curvature in cells adjacent to an IB computational cell, the authors employ a method where the IB
stencil is modified to take into account the orientation of the interface normal at the contact line to reflect the prescribed contact angle. The modified stencil 
is illustrated in Fig.~\ref{fig:IB}. At each point where the stencil intersects the IB, the prescribed contact angle defines the value of the interface normal
at the contact line, $\vec{\bf n}_{cl}$. Therefore, adjacent cells to the IB enter
into the computation of the curvature, $\kappa$, resulting in a wetting force that
consequently enters the surface tension discretization. 

\begin{center}
	\begin{figure}[t]
		\includegraphics[width=1.15\textwidth,trim=10mm 0 0 0]{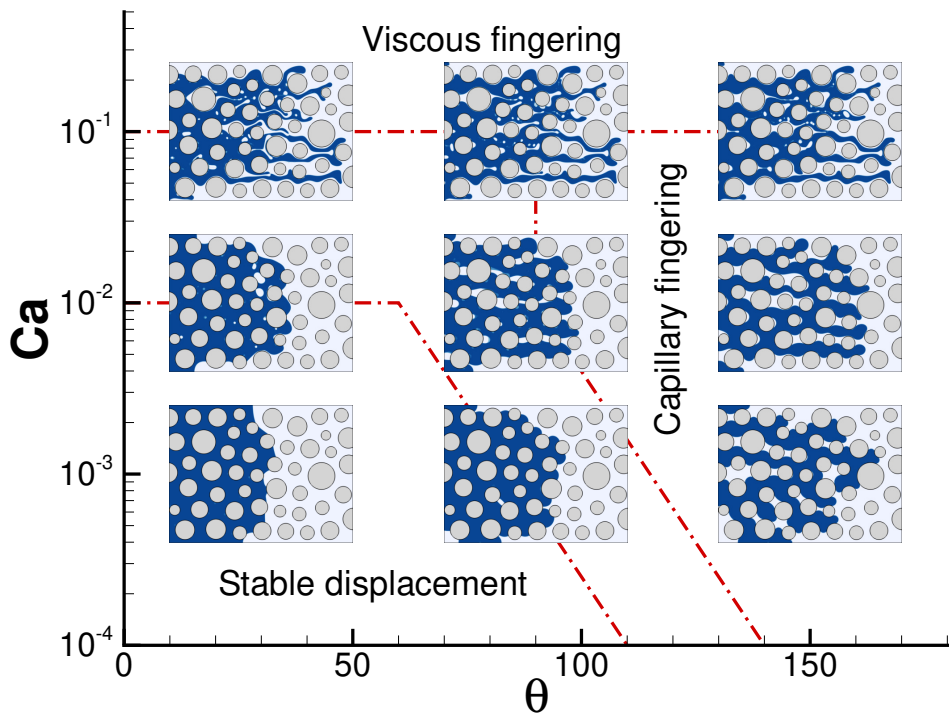}
		\caption{From \cite{OBAB2020}. Displacement patterns at varying $\Ca$ and $\theta$. Above a certain $\Ca$ threshold only viscous fingering is visible for the parameters studied in this work.}
		\label{fig:displacement_patterns}
		\vspace{-0.1in}
	\end{figure}
\end{center}
In \cite{OBAB2020}, we present the results of an improved implementation of the above method \cite{OBrien2019}
to simulate two-phase flows in a porous media model.
The results are first of their kind in that they are obtained using 
a combined VOF/IBM pore-scale direct numerical simulations to quantify the effects of the capillarity, characterized by the capillary number, $\Ca$, and wetting, characterized by the contact angle, $\theta$,
on the displacement phenomena in a porous media model. 
In Fig.~\ref{fig:displacement_patterns}, we show how wetting controls displacement patterns, and observe the crossover from a stable propagating front to a fractal pattern, namely a fingering instability. 
The results show that decreasing the contact angle leads to a change of the flow behavior from strong fingering to stabilized front propagation, and that the contact angle effect diminishes as $\Ca$ increases. As can be seen in Fig.~\ref{fig:displacement_patterns} displacement patterns,
for wetting invading fluids, the front propagates rather smoothly, draining all of the displaced phase, while a non-wetting invading fluid can lead to residual fluid.

\section{Future direction: microlayer formation in nucleate boiling as an example}
\label{sec:ML}
\begin{center}
	\begin{figure}[t]
		\includegraphics[width=1.\textwidth,trim=0 0 0 0]{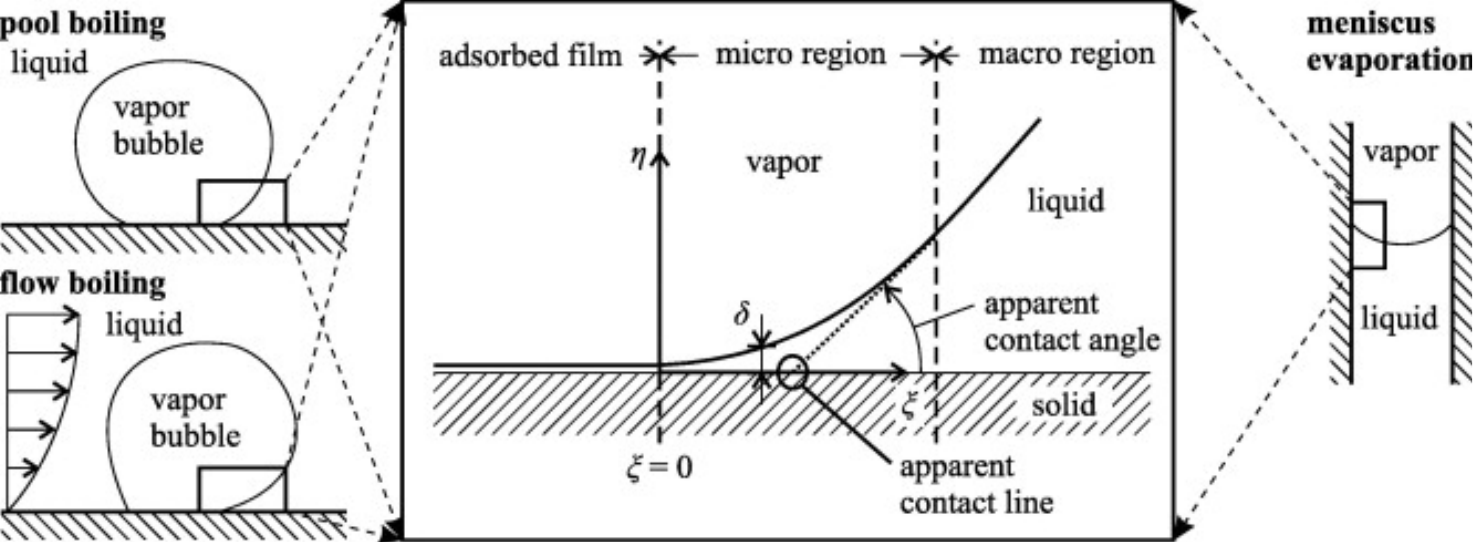}
		\caption{From \cite{KUNKELMANN2012}. Multiple scales involved in nucleate boiling: from the millimeter scale at the bubble cap to the nanometer scale at the adsorbed liquid layer. }
		\label{fig:pool}
		\vspace{-0.1in}
	\end{figure}
\end{center}

Pool boiling is one of the most efficient mode of heat transfer, 
allowing a wide range of systems to improve their thermal performance, 
from nuclear power plants to microelectronic devices \cite{Kandlikar12}.
Predicting boiling heat transfer however is complicated by the need to resolve phenomena occurring over multiple scales (see Fig.\ref{fig:pool}), from the adsorbed liquid layer at the wall at the nanometer scale up to the bubble size at the millimeter scale. Furthermore, a significant amount of the total heat can be transferred within the micro-region (known as microlayer) near the contact line.
Existing models of microlayer formation are often physically incomplete, e.g. do not include the effect of surface tension or contact angle. Moreover,
the dynamics of the moving contact line has a profound influence on
the microlayer formation. Therefore, a general model on microlayer formation
must also resolve the motion of the contact line. 

In \cite{KUNKELMANN2012}, the authors present experimental and numerical 
observations of contact line heat transfer in the framework of pool boiling and
meniscus evaporation. They focus on the influence of contact line dynamics on the
local heat transfer near the contact line suggesting that the heat transfer next to the contact line is governed mainly by two mechanisms in superposition: microlayer
evaporation and transient conduction. 
In \cite{GUION2018}, the authors present a numerical framework in which a subgrid source of vapor is considered. 
In \cite{SCHWEIKERT2019}, the authors present experimental results on the transition between contact line evaporation and microlayer evaporation during the dewetting of a superheated wall. They discuss the time resolved formation of the microlayer  in detail and describe the influence of the dewetting velocity, wall superheat, and heating power on the microlayer formation. In \cite{Sato2021}, the authors describe the microlayer formation as a dewetting transition in the presence of phase change and use the existing theoretical, experimental and numerical data to a derive a specific criterion for modeling the transition.

Here, we provide a review of the work in \cite{GUION2018} of the procedure for
direct computations of nucleate boiling and the conditions under which
the microlayer is expected to form. In \cite{GUION2018}, we focus on numerical
simulation of the hydrodynamics of bubble growth at a wall, and 
high resolution simulations of the microlayer formation. 
We identify the minimum set of dimensionless parameters that controls microlayer formation, namely
$
\delta^*= f (r^*, t^*, \Ca, \theta)
$
with $\delta^*$ the shape of the extended liquid microlayer, $r_c=\mu_l / (\rho_l U_b)$ and $t_c=r_c / U_b$ the characteristic length and time scales used, respectively, where $U_b$ is the bubble growth rate. 
We found that all three regions need to be modeled in order to accurately represent the thickness of the microlayer. Central to our theory is the results that 
the interface profile in the microlayer away from the contact line can be 
accurately described (in the case of hemispherical bubble growth) as
$
	\delta^* \approx 0.5 \sqrt {r^*- R_{b,0}^*}
$
as long as the motion of the contact line is negligible compared to the bubble growth rate;  $R_{b,0}^*$ is the dimensionless initial bubble
radius and $r^*$ is the dimensionless radial distance from the bubble root. 

\begin{figure}
	\begin{center}
		\begin{tabular}{cc}
			\includegraphics[width=.49\textwidth]{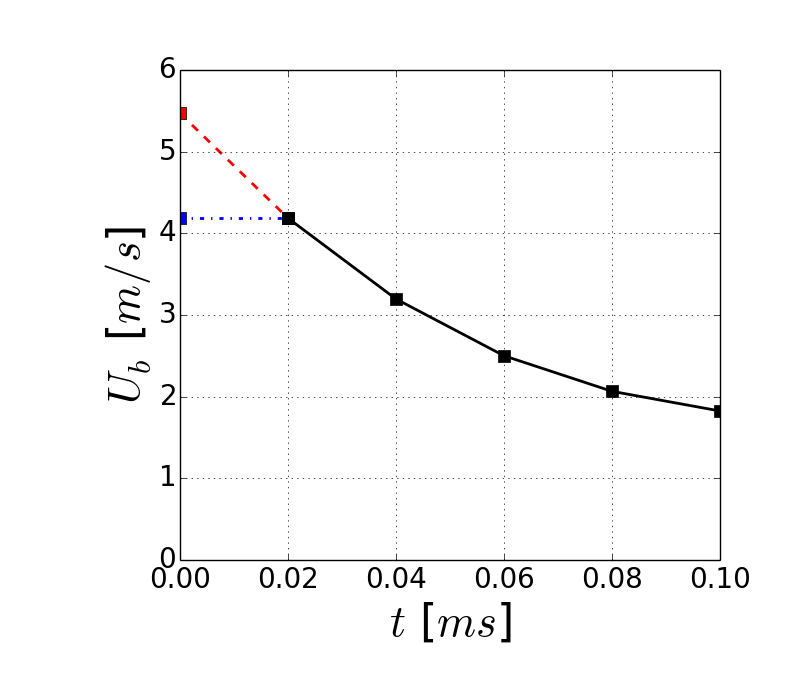}
			&\includegraphics[width=.49\textwidth]{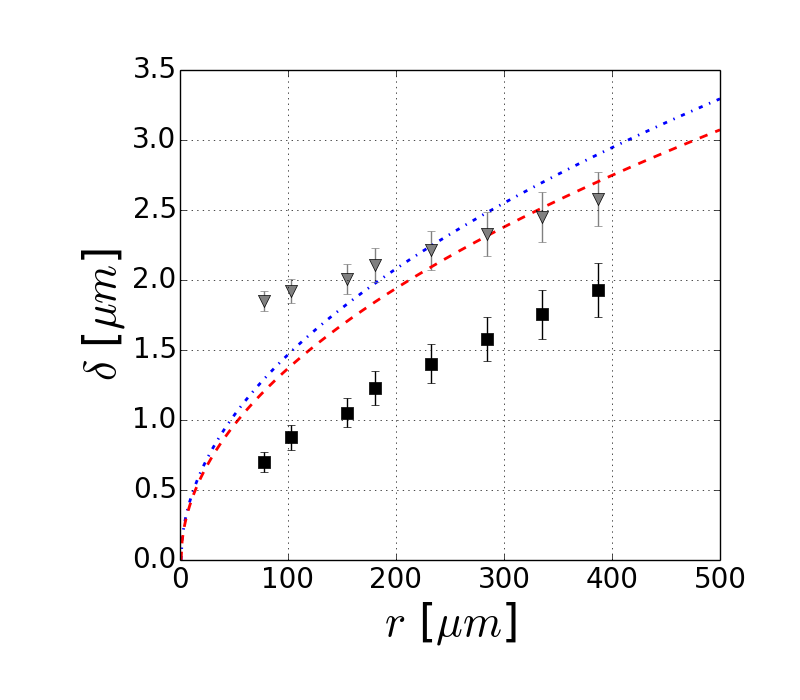}
			\\
			(a) & (b)
		\end{tabular}
		\caption{From: \cite{GUION2018}. Measured bubble growth rates (black squares) over time (a); the earliest bubble growth rate is measured at $t=20\mu s$, showing that the initial bubble growth rate during the first tens of $\mu s$ may have been much higher than its first available measured value of $4.2m/s$.
			Two scenarios assumed for initial regimes: constant growth rate (blue dash-dotted line), or constant acceleration (red dashed line). Thickness $\delta$ (in $\mu m$) measured in lab experiment (black squares) of boiling water at saturation at $0.101MPa$, $0.41ms$ after boiling inception. The deduced microlayer thickness at $t=20 \mu s$ (grey triangles) is obtained assuming that the heat transfer within the microlayer is purely by conduction. The wall temperature at boiling inception is $111.7^\circ C$, and the wall heat flux is $209kW/m^2$. Predictions from the square root model are plotted for the two initial profiles of $U_b$ showed in (a). The first measured value of $U_b$ is $U_b(t\sim20\mu s)\sim 4.2m/s$.
			For simulations: $\mu_l=3.5\times10^{-4} ~Pa.s$, $\rho_l=950 kg/m^{3}$. 
		}
		\label{benchmark}
	\end{center}
\end{figure}

Experimental measurements of the earliest bubble growth rate (Fig.~\ref{benchmark} (a))
and microlayer thickness (Fig.~\ref{benchmark} (b)) are then used in \cite{GUION2018} to validate the numerical results, for the case of water at saturation at $0.101MPa$. The numerical results are compared with experimental results in Fig.~\ref{benchmark} (b), showing both qualitative and quantitative agreements. However, a departure from the model in the low thickness region is also observed. This discrepancy is expected to be reduced if we no longer assume constant wall superheat, but include the thermal coupling between the microlayer and the heated wall. Such coupling would require to solve for the heat equation in the substrate. The initial evaporation of the microlayer would locally reduce the wall superheat and therefore slow down the rate of evaporation of the microlayer. This slow-down effect of the microlayer evaporation due to conjugate heat transfer with the solid surface is currently not included in the results in Fig.~\ref{benchmark}, and must be a subject of further investigation. In general, systematic experiments seem to be necessary to validate  the numerical model and to develop more applicable dynamic contact line models. 

\subsection{Hot topics and future issues}
Numerical challenges for dynamic contact lines amplify when including 
heat and mass transfer mechanisms (i.e.~evaporation, boiling, and condensation). 
The numerical framework described in this review will face new challenges 
to accurately and robustly implement the velocity discontinuity across
the interface due to the mass exchange. Including the conjugate heat 
transfer with the solid substrate, which requires the coupling of the
flow solver and the energy equations, as well as a domain
decomposition strategy, introduces additional challenges;
see e.g.\cite{Seric2018PoF}. Here, we list some future issues:

\begin{itemize}
	\item The continuum formulations for moving contact line models lay out a strong
	foundation for studying such systems. We believe that more elaborate models that
	include microscopic phenomena are needed to calibrate and verify numerical results 
	towards building a more robust moving contact line model.
	\item A static contact angle is not a compatible model to impose on a moving contact line.
	We believe a dynamic contact angle model should address a more physically correct model of the contact line, which is essential for simulating real applications.
	\item Hybrid numerical models that include hydrodynamic
	theories of the dynamic contact line show promise. A more general theories to large 
	contact angles are needed to extend the applicability of such hybrid models.
	Moreover, a special attention should be given on the development
	of numerical methods that include more level of details, still maintaining
	computational efficiency.
	\item Extension of all the above items to textured surfaces poses a formidable numerical 
	challenge.   
	\item This review provides an example of a powerful numerical 
	tool, Gerris, for studying relevant problems. While laying out a strong foundation,
	more improved numerical methods should continue to be developed
	for increasing the predictive power of direct simulations for real life 
	engineering applications, especially for three-dimensional domains.

\end{itemize}

\section*{Acknowledgment}

We acknowledge the support by the Petroleum Research Fund PRF-59641-ND.
We thank P.~Lehrer for help with Basilisk computations, K.~Mahady,
for fluid--fluid--solid interaction simulations, and S. Zaleski for
fruitful discussions and help with improving this review. 

\bibliography{Challenges-of-Numerical-Simulation-of-Wetting-Phenomena.bib}

\begin{thebibliography}{100}
\expandafter\ifx\csname url\endcsname\relax
  \def\url#1{\texttt{#1}}\fi
\expandafter\ifx\csname urlprefix\endcsname\relax\def\urlprefix{URL }\fi
\expandafter\ifx\csname href\endcsname\relax
  \def\href#1#2{#2} \def\path#1{#1}\fi

\bibitem{deGennesRMP}
P.~{de Gennes}, {Wetting: Statics and Dynamics}, Rev. Mod. Phys. 57 (1985) 827.

\bibitem{Jacqmin2000}
J.~Jacqmin, Contact-line dynamics of a diffuse fluid interface, J. Fluid Mech.
  402 (2000) 57.

\bibitem{Spelt2005}
P.~D. Spelt, A level-set approach for simulations of flows with multiple moving
  contact lines with hysteresis, J. Comput. Phys. 207 (2005) 389.

\bibitem{AB2008}
S.~Afkhami, M.~Bussmann, {Height functions for applying contact angles to 2{D}
  {VOF} simulations}, Int. J. Numer. Meth. Fluids 57 (2008) 453.

\bibitem{afkhami_jcp09}
S.~Afkhami, S.~Zaleski, M.~Bussmann, {A mesh-dependent model for applying
  dynamic contact angles to {VOF} simulations}, J. Comput. Phys. 228 (2009)
  5370.

\bibitem{afkhami_jcp2018}
S.~Afkhami, J.~Buongiorno, A.~Guion, S.~Popinet, Y.~Saade, R.~Scardovelli,
  S.~Zaleski, {Transition in a numerical model of contact line dynamics and
  forced dewetting}, J. Comput. Phys. 374 (2018) 1061.

\bibitem{Worner2012}
{W\"{o}rner, M.}, {Numerical modeling of multiphase flows in microfluidics and
  micro process engineering: a review of methods and applications},
  Microfluidics Nanofluidics 12 (2012) 841.

\bibitem{Sui14}
Y.~Sui, H.~Ding, P.~D.~M. Spelt, Numerical simulations of flows with moving
  contact lines, Annu. Rev. Fluid Mech. 46 (2014) 97.

\bibitem{pismen00b}
L.~M. Pismen, Y.~Pomeau, Disjoining potential and spreading of thin liquid
  layers in the diffuse-interface model coupled to hydrodynamics, Phys. Rev. E
  62 (2000) 2480.

\bibitem{POMEAU2002}
Y.~Pomeau, Recent progress in the moving contact line problem: a review,
  {Comptes Rendus M\'{e}canique} 330 (2002) 207.

\bibitem{pismen00}
L.~M. Pismen, B.~Y. Rubinstein, {Spreading of a wetting film under the action
  of van der Waals forces}, Phys. Fluids 12 (2000) 480.

\bibitem{STAROV1994}
V.~Starov, V.~Kalinin, J.-D. Chen, Spreading of liquid drops over dry surfaces,
  Adv. Colloid Interface Sci. 50 (1994) 187.

\bibitem{Bussmann99}
M.~Bussmann, J.~Mostaghimi, S.~Chandra, On a three-dimensional volume tracking
  model of droplet impact, Phys. Fluids 11 (1999) 1406.

\bibitem{Ding07}
H.~Ding, P.~D.~M. Spelt, Wetting condition in diffuse interface simulations of
  contact line motion, Phys. Rev. E 75 (2007) 046708.

\bibitem{AB2009}
S.~Afkhami, M.~Bussmann, {Height functions for applying contact angles to 3{D}
  {VOF} simulations}, Int. J. Numer. Meth. Fluids 61 (2009) 827.

\bibitem{Shin2018}
S.~Shin, J.~Chergui, D.~Juric, {Direct simulation of multiphase flows with
  modeling of dynamic interface contact angle}, Theor. Comput. Fluid Dyn. 32
  (2018) 655.

\bibitem{yue_2020}
P.~Yue, Thermodynamically consistent phase-field modelling of contact angle
  hysteresis, J. Fluid Mech. 899 (2020) A15.

\bibitem{LI2020}
H.-L. Li, H.-R. Liu, H.~Ding, A fully {3D} simulation of fluid-structure
  interaction with dynamic wetting and contact angle hysteresis, J. Comput.
  Phys. 420 (2020) 109709.

\bibitem{HAN2021}
T.-Y. Han, J.~Zhang, H.~Tan, M.-J. Ni, A consistent and parallelized height
  function based scheme for applying contact angle to 3{D} volume-of-fluid
  simulations, J. Comput. Phys. 433 (2021) 110190.

\bibitem{OBrien2019}
A.~O'Brien, M.~Bussmann, A moving immersed boundary method for simulating
  particle interactions at fluid-fluid interfaces, J. Comput. Phys. 402 (2020)
  109089.

\bibitem{popinetGerris}
S.~Popinet, {The {G}erris flow solver}, http://gfs.sourceforge.net/, 1.3.2
  (2012).

\bibitem{Popinet03}
S.~Popinet, {Gerris: a tree-based adaptive solver for the incompressible
  {E}uler equations in complex geometries}, J. Comput. Phys. 190 (2003) 572.

\bibitem{Popinet2009}
S.~Popinet, {An accurate adaptive solver for surface-tension-driven interfacial
  flows}, J. Comput. Phys. 228 (2009) 5838.

\bibitem{PopinetARFM}
S.~Popinet, Numerical models of surface tension, Ann. Rev. Fluid Mech. 50
  (2018) 49.

\bibitem{popinetBasilisk}
S.~Popinet, {{B}asilisk, a free-software program for the solution of partial
  differential equations on adaptive {C}artesian meshes}, http://basilisk.fr
  (2018).

\bibitem{Fullana2020}
T.~Fullana, S.~Zaleski, S.~Popinet, Dynamic wetting failure in curtain coating
  by the {Volume-of-Fluid} method, Eur. Phys. J. Special Topics 229 (2020)
  1923.

\bibitem{Sakakeeny2021}
J.~Sakakeeny, Y.~Ling, Numerical study of natural oscillations of supported
  drops with free and pinned contact lines, Phys. Fluids 33 (2021) 062109.

\bibitem{Blake2006}
T.~D. Blake, The physics of moving wetting lines, J. Colloid Interface Sci. 299
  (2006) 1.

\bibitem{Shikhmurzaev}
Y.~Shikhmurzaev, {Moving contact lines in liquid/liquid/solid systems}, J.
  Fluid Mech. 334 (1997) 211.

\bibitem{HuhScriv}
C.~Huh, L.~E. Scriven, {Hydrodynamic model of steady movement of a
  solid/liquid/fluid contact line}, J. Colloid Interface Sci. 35 (1971) 85.

\bibitem{Dziedzic2019}
A.~Dziedzic, M.~Nakrani, B.~Ezra, M.~Syed, S.~Popinet, S.~Afkhami, Breakup of
  finite-size liquid filaments: {T}ransition from no-breakup to breakup
  including substrate effects, Eur. Phys. J. E 42 (2019) 18.

\bibitem{LIU05}
H.~Liu, S.~Krishnan, S.~Marella, H.~Udaykumar, {Sharp interface Cartesian grid
  method II: A technique for simulating droplet interactions with surfaces of
  arbitrary shape}, J. Comput. Phys. 210 (2005) 32.

\bibitem{ZHANG2020}
J.~Zhang, P.~Yue, A level-set method for moving contact lines with contact
  angle hysteresis, J. Comput. Phys. 418 (2020) 109636.

\bibitem{cahn_jcp58}
J.~W. Cahn, J.~E. Hilliard, {Free energy of a nonuniform system. 1.
  {I}nterfacial free energy}, J. Chem. Phys. 28 (1958) 258.

\bibitem{Jacqmin1999}
J.~Jacqmin, Calculation of two-phase navier-stokes flows using phase field
  modeling, J. Comp. Phys. 155 (1999) 96.

\bibitem{Jacqmin2004}
J.~Jacqmin, Onset of wetting failure in liquid-liquid systems, J. Fluid Mech.
  517 (2004) 209.

\bibitem{Yue2010}
P.~Yue, C.~Zhou, J.~Feng, Sharp-interface limit of the {C}ahn-{H}illiard model
  for moving contact lines, J. Fluid Mech. 645 (2010) 279.

\bibitem{Sibley2013}
D.~Sibley, A.~Nold, S.~Kalliadasis, Unifying binary fluid diffuse-interface
  models in the sharp interface limit, J. Fluid Mech. 736 (2013) 5.

\bibitem{Briant2004}
A.~J. Briant, A.~J. Wagner, J.~M. Yeomans, Lattice boltzmann simulations of
  contact line motion. i. liquid-gas systems, Phys. Rev. E. 69 (2004) 031602.

\bibitem{Lee2010}
T.~Lee, L.~Liu, Lattice boltzmann simulations of micron-scale drop impact on
  dry surfaces, J. Comp. Phys. 229 (2010) 8045.

\bibitem{Qian2003}
T.~Qian, X.-P. Wang, P.~Sheng, Molecular scale contact line hydrodynamics of
  immiscible flows, Phys. Rev. E 68 (2003) 016306.

\bibitem{Qian2006}
T.~Qian, X.-P. Wang, P.~Sheng, Molecular hydrodynamics of the moving contact
  line in two-phase immiscible flows, Comm. Comput. Phys. 1 (2006) 1.

\bibitem{Nguyen2012}
T.~Nguyen, M.~Fuentes-Cabrera, J.~Fowlkes, J.~Diez, A.~Gonz{\'a}lez, L.~Kondic,
  P.~Rack, Competition between collapse and breakup in nanometer-sized thin
  rings using molecular dynamics and continuum modeling, Langmuir 28 (2012)
  13960.

\bibitem{fuentes_pre11}
M.~Fuentes-Cabrera, B.~Rhodes, J.~Fowlkes, A.~L{\'o}pez-Benzanilla,
  H.~Terrones, M.~Simpson, P.~Rack, {Molecular dynamics study of the dewetting
  of copper on graphite and graphene: Implications for nanoscale
  self-assembly}, Phys. Rev. E 83 (2011) 041603.

\bibitem{afkhami-kondic-2013}
S.~Afkhami, L.~Kondic, Numerical simulation of ejected molten metal
  nanoparticles liquified by laser irradiation: Interplay of geometry and
  dewetting, Phys. Rev. Lett. 111 (2013) 034501.

\bibitem{Ugis2020}
{U. L\v{a}cis, P. Johansson, T. Fullana, B. Hess, G. Amberg, S. Bagheri, and S.
  Zaleski}, Steady moving contact line of water over a no-slip substrate, Eur.
  Phys. J. Special Topics 229 (2020) 1897.

\bibitem{REN2005}
W.~Ren, W.~E, Heterogeneous multiscale method for the modeling of complex
  fluids and micro-fluidics, J. Comput. Phys. 204 (2005) 1.

\bibitem{Huang2004}
H.~Huang, D.~Liang, B.~B.~Wetton, {Computation of a moving drop/bubble on a
  slid surface using a front-tracking method}, Commun. Math. Sci. 2 (2004) 535.

\bibitem{MahadyvdW15}
K.~Mahady, S.~Afkhami, L.~Kondic, A volume of fluid method for simulating
  fluid/fluid interfaces in contact with solid boundaries, J. Comp. Phys. 294
  (2015) 243.

\bibitem{Brackbill92}
J.~U. Brackbill, D.~B. Kothe, C.~Zemach, {A continuum method for modeling
  surface tension}, J. Comput. Phys. 100 (1992) 335.

\bibitem{MahadyvdW2015}
K.~Mahady, S.~Afkhami, L.~Kondic, A numerical approach for the direct
  computation of flows including fluid-solid interaction: modeling contact
  angle, film rupture, and dewetting, Phys. Fluids 28 (2016) 062002.

\bibitem{cox1986}
R.~G. Cox, {The dynamics of the spreading of liquids on a solid surface. Part
  1. Viscous flow}, {J. Fluid Mech.} 168 (1986) 169.

\bibitem{voinov1976}
O.~Voinov, {Hydrodynamics of wetting}, Fluid Dynamics 11 (1976) 714.

\bibitem{bonn_rmp2009}
D.~Bonn, J.~Eggers, J.~Indekeu, J.~Meunier, E.~Rolley, {Wetting and spreading},
  Rev. Mod. Phys. 81 (2009) 739.

\bibitem{seeman_jphys01}
R.~Seemann, S.~Herminghaus, K.~Jacobs, {Gaining control of pattern formation of
  dewetting liquid films}, J. Phys. Condens. Matt. 21 (2001) 4925.

\bibitem{neto_jphys03}
C.~Neto, K.~Jacobs, R.~Seemann, R.~Blossey, J.~Becker, G.~Gr{\"u}n, {Satellite
  hole formation during dewetting: experiment and simulation}, J. Phys.:
  Condens. Matter 15 (2003) 3355.

\bibitem{becker_nat03}
J.~Becker, G.~Gr{\"u}n, R.~Seemann, H.~Mantz, K.~Jacobs, K.~R. Mecke,
  R.~Blossey, {Complex dewetting scenarios captured by thin-film models},
  Nature Mat. 2 (2003) 59.

\bibitem{OH2018}
Y.~Oh, J.~Lee, M.~Lee, Fabrication of {A}g-{A}u bimetallic nanoparticles by
  laser-induced dewetting of bilayer films, Appl. Surf. Sci. 434 (2018) 1293.

\bibitem{Rack2020}
D.~A. Garfinkel, G.~Pakeltis, N.~Tang, I.~N. Ivanov, J.~D. Fowlkes, D.~A. G.,
  P.~D. Rack, Optical and magnetic properties of ag–ni bimetallic
  nanoparticles assembled via pulsed laser-induced dewetting, ACS Omega 5
  (2020) 19285.

\bibitem{Fricke2020}
{M. Fricke, D. Bothe}, {Boundary conditions for dynamic wetting - A
  mathematical analysis}, Eur. Phys. J. Special Topics 229 (2020) 1849.

\bibitem{AGRP}
S.~Afkhami, T.~Gambaryan-Roisman, L.~M. Pismen, Challenges in nanoscale physics
  of wetting phenomena, Eur. Phys. J. Special Topics 229 (2020) 1735.

\bibitem{Snoeijer13}
J.~H. Snoeijer, B.~Andreotti, Moving contact lines: Scales, regimes, and
  dynamical transitions, Annu. Rev. Fluid Mech. 45 (2013) 269.

\bibitem{Tryggvason11}
G.~Tryggvason, R.~Scardovelli, S.~Zaleski, Direct Numerical Simulations of
  Gas-Liquid Multiphase Flows, Cambridge University Press, 2011.

\bibitem{LandauLevich1942}
L.~D. Landau, B.~V. Levich, Dragging of a liquid by a moving plate, Acta
  Physicochim. URSS 17 (1942) 42.

\bibitem{Derjaguin1943}
B.~V. Derjaguin, On the thickness of a layer of liquid remaining on the walls
  of vessels after their emptying, and the theory of the application of
  photoemulsion after coating on the cine film, Acta Physicochim. URSS 20
  (1943) 349.

\bibitem{Eggers2004b}
J.~Eggers, Hydrodynamic theory of forced dewetting, Phys. Rev. Lett. 93 (2004)
  094502.

\bibitem{Eggers2005}
J.~Eggers, {Contact line motion for partially wetting fluids}, {Phys. Rev. E}
  72 (2005) 061605.

\bibitem{Chan12}
T.~S. Chan, J.~H. Snoeijer, J.~Eggers, Theory of the forced wetting transition,
  Phys. Fluids 24 (2012) 072104.

\bibitem{Legendre2015}
D.~Legendre, M.~Maglio, Comparison between numerical models for the simulation
  of moving contact lines, Comput. Fluids 113 (2015) 2.

\bibitem{afkhami_jcp2018_corr}
S.~Afkhami, J.~Buongiorno, A.~Guion, S.~Popinet, Y.~Saade, R.~Scardovelli,
  S.~Zaleski, {Corrigendum to ``Transition in a numerical model of contact line
  dynamics and forced dewetting'' [J.~Comput.~Phys.~374(2018)1061--1093]}, J.
  Comput. Phys. 382 (2019) 61.

\bibitem{Qin:18at}
J.~Qin, P.~Gao, Asymptotic theory of fluid entrainment in dip coating, J. Fluid
  Mech. 844 (2018) 1026.

\bibitem{Moriarty92}
J.~A. Moriarty, L.~W. Schwartz, Effective slip in numerical calculations of
  moving-contact-line problems, J. Eng. Math. 26 (1992) 81.

\bibitem{Weinstein08}
O.~Weinstein, L.~M. Pismen, Scale dependence of contact line computations,
  Math. Model. Nat. Phenom. 3 (2008) 98.

\bibitem{YueFeng2011}
P.~Yue, J.~J. Feng, Wall energy relaxation in the {C}ahn-{H}illiard model for
  moving contact lines, Phys. Fluids 23 (2011) 012106.

\bibitem{blake1999experimental}
T.~Blake, M.~Bracke, Y.~Shikhmurzaev, Experimental evidence of nonlocal
  hydrodynamic influence on the dynamic contact angle, Phys. Fluids 11 (1999)
  1995.

\bibitem{Wilson:2006ks}
M.~Wilson, J.~Summers, Y.~Shikhmurzaev, A.~Clarke, T.~Blake, {Nonlocal
  hydrodynamic influence on the dynamic contact angle: Slip models versus
  experiment}, Phys. Rev. E 73 (2006) 041606.

\bibitem{Qian03}
T.~Qian, X.-P. Wang, P.~Sheng, {Generalized {N}avier boundary condition for the
  moving contact line}, Commun. Math. Sci. 1 (2003) 333.

\bibitem{FRICKE2019}
M.~Fricke, M.~K\"{o}hne, D.~Bothe, A kinematic evolution equation for the
  dynamic contact angle and some consequences, Phys. D: Nonlinear Phenom. 394
  (2019) 26.

\bibitem{Wang2008}
X.-P. Wang, T.~Qian, P.~Sheng, Moving contact line on chemically patterned
  surfaces, J. Fluid Mech. 605 (2008) 59.

\bibitem{Xu2011}
X.~Xu, X.-P. Wang, Analysis of wetting and contact angle hysteresis on
  chemically patterned surfaces, SIAM J. Appl. Math. 71 (2011) 1753.

\bibitem{Ren2014}
W.~Ren, Wetting transition on patterned surfaces: {T}ransition states and
  energy barriers, Langmuir 30 (2014) 2879.

\bibitem{LUO2017}
L.~Luo, X.-P. Wang, X.-C. Cai, An efficient finite element method for
  simulation of droplet spreading on a topologically rough surface, J. Comput.
  Phys. 349 (2017) 233.

\bibitem{Li2021}
W.~Li, J.~Lu, G.~Tryggvason, Y.~Zhang, Numerical study of droplet motion on
  discontinuous wetting gradient surface with rough strip, Phys. Fluids 33
  (2021) 012111.

\bibitem{Ferrari2013}
A.~Ferrari, I.~Lunati, Direct numerical simulations of interface dynamics to
  link capillary pressure and total surface energy, Adv. Water Resour. 57
  (2013) 19.

\bibitem{Raeini2014}
A.~Q. Raeini, M.~J. Blunt, B.~Bijeljic, Direct simulations of two-phase flow on
  micro-ct images of porous media and upscaling of pore-scale forces, Adv.
  Water Resour. 57 (2014) 116.

\bibitem{Bakhshian2019}
S.~Bakhshian, S.~Hosseini, N.~Shokri, Pore-scale characteristics of multiphase
  flow in heterogeneous porous media using the lattice {B}oltzmann method, Adv.
  Water Resour. 9 (2019) 3377.

\bibitem{Zhao2019}
B.~Zhao, C.~W. MacMinn, B.~K. e.~a. Primkulov, Comprehensive comparison of
  pore-scale models for multiphase flow in porous media, Proc. Natl. Acad. Sci.
  116~(28) (2019) 13799.

\bibitem{Ghillani2020}
{N. Ghillani, M. Heinz, T. Gambaryan-Roisman}, Capillary rise and evaporation
  of a liquid in a corner between a plane and a cylinder: A model of imbibition
  into a nanofiber mat coating, Eur. Phys. J. Special Topics 229 (2020) 1799.

\bibitem{BASIRAT2017}
F.~Basirat, Z.~Yang, A.~Niemi, {Pore-scale modeling of wettability effects on
  CO$_2$--brine displacement during geological storage}, Adv. Water Resour. 109
  (2017) 181.

\bibitem{LIU2014}
H.~Liu, A.~J. Valocchi, C.~Werth, Q.~Kang, M.~Oostrom, {Pore-scale simulation
  of liquid CO$_2$ displacement of water using a two-phase lattice Boltzmann
  model}, Adv. Water Resour. 73 (2014) 144.

\bibitem{Peskin1972}
C.~S. Peskin, Flow patterns around heart valves: {A} numerical method, J.
  Comput. Phys. 10 (1972) 252.

\bibitem{Mittal2005}
R.~Mittal, G.~Iaccarino, {Immersed Boundary Methods}, Annu. Rev. Fluid Mech. 37
  (2005) 239.

\bibitem{OBrien2018}
A.~O'Brien, M.~Bussmann, A volume-of-fluid ghost-cell immersed boundary method
  for multiphase flows with contact line dynamics, Comput. {\&} Fluids 165
  (2018) 43.

\bibitem{OBAB2020}
{A. O'Brien and S. Afkhami and M. Bussmann}, {Pore-scale direct numerical
  simulation of Haines jumps in a porous media model}, Eur. Phys. J. Special
  Topics 229 (2020) 1785.

\bibitem{Kandlikar12}
S.~G. Kandlikar, {History, Advances, and Challenges in Liquid Flow and Flow
  Boiling Heat Transfer in Microchannels: A Critical Review}, J. Heat Transfer
  134.

\bibitem{KUNKELMANN2012}
C.~Kunkelmann, K.~Ibrahem, N.~Schweizer, S.~Herbert, P.~Stephan,
  T.~Gambaryan-Roisman, {The effect of three-phase contact line speed on local
  evaporative heat transfer: Experimental and numerical investigations}, Int.
  J. Heat Mass Transfer 55 (2012) 1896.

\bibitem{GUION2018}
A.~Guion, S.~Afkhami, S.~Zaleski, J.~Buongiorno, Simulations of microlayer
  formation in nucleate boiling, Int. J. Heat Mass Transfer 127 (2018) 1271.

\bibitem{SCHWEIKERT2019}
K.~Schweikert, A.~Sielaff, P.~Stephan, On the transition between contact line
  evaporation and microlayer evaporation during the dewetting of a superheated
  wall, Int. J. Thermal Sci. 145 (2019) 106025.

\bibitem{Sato2021}
L.~Bure\v{s}, Y.~Sato, On the modelling of the transition between contact-line
  and microlayer evaporation regimes in nucleate boiling, J. Fluid Mech. 916
  (2021) A53.

\bibitem{Seric2018PoF}
I.~Seric, S.~Afkhami, L.~Kondic, Influence of thermal effects on stability of
  nanoscale films and filaments on thermally conductive substrates, Phys.
  Fluids 30 (2018) 012109.

\end{thebibliography}

\end{document}